\documentstyle[12pt,epsf]{article}
\newcommand\pubnumber{SLAC-PUB-8319}
\newcommand\stanfordnumber{SU-ITP-99/53}
\newcommand\pubdate{January, 2000}

\def\Title#1{\begin{center} {\Large #1 } \end{center}}
\def\Author#1{\begin{center}{ \sc #1} \end{center}}
\def\Address#1{\begin{center}{ \it #1} \end{center}}

\def\submit#1{\begin{center}Submitted to {\sl #1} \end{center}}
\def\doeack{\footnote{Work supported by the Department of Energy,
                     contract DE--AC03--76SF00515.}}
\def\stanfordack{\footnote{Work supported in part by the National Science
              Foundation,
                     contract PHY--9870115.}}
\def\SLAC{Stanford Linear Accelerator Center\\
    Stanford University, Stanford, California 94309 USA}
\def\Stanford{Department of Physics\\
    Stanford University, Stanford, California 94305 USA}
\newcommand\pubblock{\rightline{\begin{tabular}{l} \pubnumber\\
               \stanfordnumber \\
         \pubdate  \end{tabular}}}
\newenvironment{Abstract}{\begin{quotation} \begin{center}
                       ABSTRACT
     \end{center}\bigskip  }{\end{quotation}}
\def\beq{\begin{equation}}
\def\eeq#1{\label{#1}\end{equation}}
\def\eeqn{\end{equation}}
\def\beqa{\begin{eqnarray}}
\def\eeqa#1{\label{#1}\end{eqnarray}}
\def\eeqan{\end{eqnarray}}
\def\CR{\nonumber \\ }
\def\leqn#1{(\ref{#1})}
\def\Acknowledgements{\bigskip  \bigskip \begin{center} \begin{large}
             \bf ACKNOWLEDGEMENTS \end{large}\end{center}}
\let\bar=\overbar
\def\Dslash{\not{\hbox{\kern-4pt $D$}}}
\def\dslash{\not{\hbox{\kern-2pt $\del$}}}
\def\half{\frac{1}{2}}

\def\A{{\cal A}}

\def\F{{\cal F}}
\def\L{{\cal L}}
\def\M{{\cal M}}

\def\V{{\cal V}}
\def\S{{\cal S}}

\def\tr{{\mbox{\rm tr}}}
\def\del{\partial}

\def\Pl{{\mbox{\scriptsize Pl}}}

\def\CM{{\mbox{\scriptsize CM}}}
\def\SM{{\mbox{\scriptsize SM}}}

\def\ee{e^+e^-}
\def\sstw{\sin^2\theta_w}
\def\cstw{\cos^2\theta_w}
\def\mz{m_Z}

\def\msb{{\bar{\ssstyle M \kern -1pt S}}}

\def\ELER{e^-_Le^+_R}

\def\s#1{\widetilde{#1}}
\def\etal{{\it et al.}}

\def\ap{{\alpha^\prime}}

\def\G{\Gamma}

\def\eps{\epsilon}
\def\V{{\cal V}}
\def\th{\vartheta}

\begin{document}
\begin{titlepage}
\pubblock

\vfill
\Title{TeV Strings and Collider Probes of Large Extra Dimensions}
\vfill
\Author{Schuyler Cullen\stanfordack}
\medskip
\Address{\Stanford}
\medskip
\medskip
\Author{Maxim Perelstein and Michael E. Peskin\doeack}
\medskip
\Address{\SLAC}
\vfill
\begin{Abstract}
Arkani-Hamed, Dimopoulos, and Dvali have proposed that the fundamental
gravitational scale is close to 1 TeV, and that the observed weakness
of gravity at long distances is explained by the presence of
large extra compact dimensions.  If this scenario is realized in a string
theory of quantum gravity, the string excited states of Standard Model
particles will also have TeV masses.  These states will be visible to
experiment and in fact provide the first signatures of the presence of
a low quantum gravity scale.
Their presence also affects the more familiar signatures due to
real and virtual graviton emission.  We study the effects of these
states in a simple string model.
\end{Abstract}
\medskip
\submit{Physical Review {\bf D}}

\vfill
\end{titlepage}
\def\thefootnote{\fnsymbol{footnote}}
\setcounter{footnote}{0}
\section{Introduction}

Traditionally, the weakness of gravitational interactions at the
scales accessible to particle physics experiments has been explained
by postulating that the Planck scale at which gravity becomes strong
 is very high, $M_\Pl \sim 10^{19}$ GeV.  Below this
scale, ordinary quantum field theory applies, but, when this scale is
reached, one can observe the underlying quantum theory that incorporates
quantum gravity.  A disappointing feature of the traditional
framework is that the enormously high value of the Planck scale
prevents us from observing any effects of quantum gravity in
laboratory experiments in the conceivable future, which means that the
search for the quantum theory of gravity has to proceed without any
experimental input. Recently Arkani-Hamed, Dimopoulos, and
Dvali  (ADD) \cite{Nima1}
have proposed an alternative to this pessimistic scenario.  They have
constructed models in which gravity becomes strong at a scale $M$ of
order 1 TeV.  They explain the apparent weakness of gravity
at lower energies by the
presence of compact dimensions with compactification radius $R \gg
M^{-1}$.  We will call these `large extra dimensions'.
In this framework, gravity could have significant effects on particle
interactions at the energies accessible to current experiments and
observations \cite{Nima2}.

So far, almost all work on the phenomenological implications of large extra
dimensions has concentrated on the effects of real and virtual graviton
emission. It is the basic assumption of the model that gravitons can move
in the extra dimensions.  Then the graviton quantum states will be
characterized by a (quantized) momentum in the extra dimensions. The states
with nonzero momentum are   called Kaluza-Klein (KK) excitations; they
can be described equivalently as massive spin-2 particles in
4 dimensions, with mass equal to the higher-dimensional momentum, which
couple to Standard Model particles through a coupling
to the energy-momentum tensor $T^{\mu\nu}$
with strength $M_\Pl^{-1}$. The sum over these states leads
gravity to become strong at a scale $M \ll M_\Pl$ because the  spectrum of
KK excitations becomes exceedingly dense as the size $R$ of the compact
dimensions is taken to be  much larger than $M^{-1}$.

Because the low-energy coupling of the KK excitations is model-independent,
one can study processes in which gravitons are emitted into the extra
dimensions \cite{Jim,MPP,us} in the context of a low-energy effective
field theory.  For collision energies much less than $M$,
 the cross sections for missing-energy signatures are not sensitive to the
details of  physics at the scale $M$.  This fact allows one to obtain
model-independent bounds on $M$.  On the other hand, it means that
the simple observation of  graviton emission does not give information about
the nature of the fundamental gravity theory.

The approach of low-energy effective field theory can also be applied to
processes in which the KK excitations appear as virtual exchanges contributing
to the scattering of Standard Model particles \cite{Jim,Hewett,HLZ}.
In this case, the contribution of low-energy effective field theory is
cutoff-dependent and of the same order as that from possible higher-dimension
operators.  In phenomenological analyses, the virtual KK exchange
is typically represented as a dimension-8 contact interaction of the
form $T^{\mu\nu}T_{\mu\nu}$ with a coefficient proportional to $1/M^4$.
The precise value of this coefficient depends on the underlying model.
It is also possible that this model could predict additional contact
interactions with a different spin structure that could also be observed
as corrections to Standard Model scattering processes.  For these reasons,
the virtual exchanges cannot be used to put lower bounds on $M$.  On the
other hand, the presence of high-spin contact interactions can produce
impressive signals, and the measurement of the coefficients of these
interactions can give new information on the fundamental theory.

The study of large extra dimensions differs from other phenomenological
problems in that the underlying theory from which the low-energy effective
description is derived is a theory of quantum gravity.  This fact may bring
in new and unforseen consequences.  In particular, the only known framework
that allows a self-consistent description of quantum gravity is string
theory \cite{firststring}.  But string theory is not simply a theory of
quantum gravity.  As an essential part of its structure, not only the gravitons
but also the particles of the Standard Model must have an extended structure.
This means that, in a string theory description, there will be additional
modifications of Standard Model amplitudes due to string excitations which
might compete with or even overwhelm the modifications due to graviton
exchange.

In this paper, we will study the signatures of string theory in a simple
toy model with large extra dimensions.  The most
important effects in this model come from the exchange of string Regge (SR)
excitations
of Standard Model particles.  We will show that, in Standard Model
scattering processes, contact interactions due to
SR exchange produce their own characteristic effects in differential
cross sections.  We will also show that these typically dominate the
effects due to KK exchange.  In addition, the SR excitations can be
directly produced as resonances.
These effects have been discussed previously, but at a more qualitative level,
by Lykken \cite{Lyk}, and by Tye and collaborators \cite{Tye}.
The effects of SR resonances have
also been studied some time ago, in the context of composite models of
quarks and leptons, by Bars and Hinchliffe \cite{BH}.

\begin{figure}
\centerline{\epsfysize=3.30truein \epsfbox{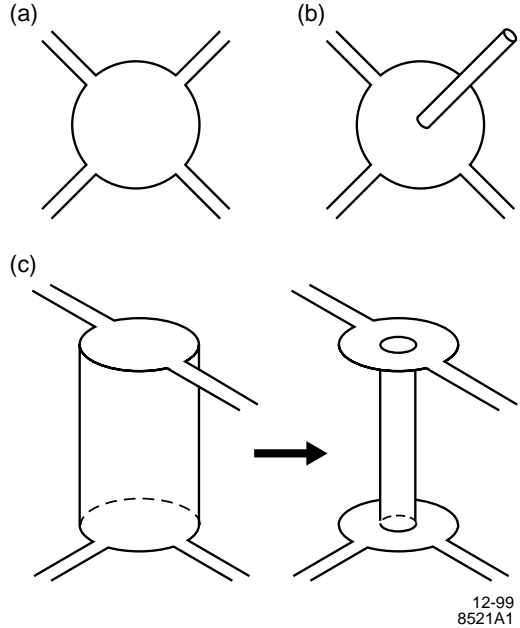}}
\vskip 0.0 cm
 \caption{Schematic diagrams contributing to scattering amplitudes in
        a string generalization of the Standard Model: (a) tree-level 2-body
        scattering; (b) graviton emission; (c) loop-level 2-body scattering.}
\label{fig:counting}
\end{figure}

The dominance of SR over KK
effects is a generic feature of weakly-coupled string
theory.  It follows from the counting of coupling constants in string
perturbation theory \cite{bigbook}, which is illustrated in
Figure~\ref{fig:counting}.
To model the ADD scenario, we consider open
string theories which contain at low energy a set of Yang-Mills
gauge bosons that can be identified with gauge bosons of the Standard Model.
We denote the dimensionless Yang-Mills coupling by~$g$.
Figure~\ref{fig:counting}(a) shows the string generalization of a Standard
Model two-body scattering amplitude at order $g^2$.  This
amplitude coincides with the Standard Model expectation in the limit in which
the center-of-mass energy is much lower than the string scale $M_S$ and,
at higher energy, shows corrections proportional to powers of  $(s/M_S^2)$.
These are the effects of SR excitations.  Figure~\ref{fig:counting}(b)
shows the leading string contribution to graviton emission.  The graviton
is a closed string state, and thus this process involves the closed-string
coupling constant, which is of order $g^2$; the full amplitude is of order
$g^3$.  Figure~\ref{fig:counting}(c) shows one contribution to the one-loop
corrections to two-body scattering.  This diagram is of order $g^4$.  However,
as Lovelace \cite{Lovelace} originally showed,
this string diagram contains the
graviton-exchange contribution when factorized as indicated in the figure.
Thus, the exchange of gravitons and their KK excitations are suppressed with
respect to SR exchange by a factor $g^2$ in the amplitude.

%%%%%%%%%%%%%%%%%%%%%%%%%%%%%%%%%%%%%%%%%%%%%%%%%%%%%%%%%%%%%%%%%%%%%%
%%%%%%%%%%%%%%%%%%%%%%%%%%%%%%%%%%%%%%%%%%%%%%%%%%%%%%%%%%%%%%%%%%%%%%

In this paper, we will flesh out the picture represented by
Figure~\ref{fig:counting} using an illustrative  toy string model.
In Section 2, we will present this model, which uses scattering amplitudes
on the 3-brane of weakly-coupled Type IIB string theory
to describe a string version
of Quantum Electrodynamics with electrons and photons.  In Section 3,
we will apply this model to compute the cross sections for
Bhabha scattering and $\ee \to \gamma\gamma$ at
high energy.  In Section 4, we will
discuss the phenomenological consequences of
those results, both for contact interactions in high-energy scattering
and for the direct observability of SR resonances.  We will find a
direct bound on the string scale of $M_S > 1$ TeV.  Translated into
a bound on the fundamental quantum gravity scale, this becomes $M > 1.6$ TeV.
This bound is admittedly model-dependent, but it is also larger than any
other current limit by more than a factor of two for the relevant case of
6 large extra dimensions.

In the remainder of the paper, we will discuss the more familiar
signatures of large extra dimensions in string language.
In Section 5, we will study the KK graviton emission process
$\ee\to \gamma G$.  In Section 6, we will discuss the effects of virtual
KK graviton exchange through a detailed analysis of the process of
$\gamma\gamma$ elastic scattering.  This analysis will also allow us to
derive the relation between the string scale and the fundamental quantum
gravity scale.
In Section 7, we will review the
collider limits on large extra dimensions in the
light of the new picture presented in this paper.
Section 8 will present our conclusions.   A series of appendices review
formulae for the analysis of Bhabha scattering and present some of the more
technical details of the string calculations.

A number of the topics considered in Sections 5 and 6 have recently been
considered, from a slightly different point of view, in a paper of
Dudas and Mourad \cite{DM}.  The phenomenological importance of SR resonances
in models with a low string scale has been discussed briefly by
Accomando, Antoniadis, and Benakli \cite{AAB}.

\section{The model}

In this paper, we would like to investigate the simplest model that illustrates
the influence of string Regge (SR) excitations on physical cross sections.
Thus, we will be content to study a simple embedding of the Quantum
Electrodynamics of electrons and photons into string theory.  This theory
contains only one gauge group and only vectorlike couplings.
  More realistic string models with large extra dimensions
have been constructed by Kakuzhadze, Tye, and Shiu \cite{SKT},
Antoniadis, Bachas, and Dudas \cite{ABD}, and Ibanez, Rabadan, and
Uranga \cite{IRU}.  These
models are quite complicated. The added structure is inessential to the
general phenomenological picture that we will present in this paper, though
there are many model-dependent details that would be interesting to study.

With this motivation, we consider a very simple embedding of QED into
Type IIB string theory.
In this theory, there exists a stable BPS object, the D3-brane,
which is a 4-dimensional hypersurface on which open
strings may end. We will assume that the 10-dimensional space of the
theory has 6 dimensions
compactified on a torus with a periodicity $2\pi R$, and that $N$ coincident
D3-branes are stretched out in the 4 extended dimensions.
  The massless
states associated with open strings that end on the branes are
described by an $N=4$ supersymmetric Yang-Mills theory with a gauge
group $U(N)$.  These states include
gauge bosons $A^{\mu a}$, gauginos $\s g^{ai}$, and complex scalars
$\phi^a$, where $a$ is an index of
the adjoint representation of $U(N)$ and $i$ runs from 1 to 4.  We will
project this theory down to a $U(1)$ gauge theory with two massless
Weyl fermions and identify the gauge boson and fermions of that theory
with the photon and electron of QED.

 We take the  parameters of this theory to be the string
scale  $M_S =
{\ap}^{-1/2}$ and the (dimensionless)
Yang-Mills  coupling constant $g$, which we identify with a Standard Model
gauge coupling.  (Except for this definition of $g$,
 we adopt the conventions of \cite{bigbook}).
 Note that $M_S$ is directly observable: The SR resonances occur
at masses $M_n = \sqrt{n} M_S$, for $n = 1,2,\ldots$.

The gravitational
constant and other physical scales in the theory are derivable from
$M_S$ and $g$.  However, the relation involves one-loop calculations and
is model-dependent, depending on the full spectrum of the theory.
Quite generally in the  ADD scenario, the Newton constant which represents
the observed strength of gravity  is given in terms of the fundamental
gravitational scale $M$ by the relation \cite{Nima2,Jim,MPP,butnote}
\beq
(4 \pi G_N)^{-1} = M^{n+2} R^n \ ,
\eeq{fundamental}
where the compact dimensions are taken to be flat and periodic with
period $2\pi R$.  Our toy model corresponds to the case $n=6$.
In Section 6  we will present a simple but model-dependent computation
of the relation between $M$ and string scale
$M_S$.  We will show that
\beq
{M \over M_S} = \left({1\over \pi}\right)^{1/8} \alpha^{-1/4} \ ,
\eeq{MMSrelation}
 where $\alpha = g^2/4\pi$.  Then, for two extreme choices,
\beqa
    \alpha = 1/137  &\to&  M/M_S = 3.0  \ ; \CR
    \alpha = \alpha_s(\mbox{1\ TeV}) &\to &  M/M_S = 1.6 \ .
\eeqa{MMS}
In scattering amplitudes
involving virtual gravitons, the gravity scale will enter as $M^{-4}$, and so
the string and gravity effects will be well-separated in size.
For future reference,
the tension of the
D3-brane is given by \cite{bigbook}
\beq
 \tau_3 = {1 \over 8 \pi^3} \alpha^{-1} M_S^4 \ .
\eeq{tension}

The relations in \leqn{MMS} illustrate the most problematical aspect of our
analysis.  The naive string constructions we will use in this paper require
all of the Standard Model gauge couplings to be unified at the string scale.
Proposals for splitting these couplings to realistic values using the
vacuum expectation value of a string modulus field are given
in \cite{ABD,IRU}.  However, in this paper we will deal with the Standard
Model interactions only one at a time.

The explicit embedding that we will use is the following:  Consider the
$SU(2)$ subgroup of $U(N)$ with generators
\beq
       t^+ = {1\over \sqrt{2}} \pmatrix{0 & 1 \cr 0 & 0\cr} \ , \qquad
       t^- = {1\over \sqrt{2}} \pmatrix{0 & 0 \cr 1 & 0\cr} \ , \qquad
       t^3 = {1\over 2} \pmatrix{1 & 0 \cr 0 & -1\cr} \ .
\eeq{threelambdas}
(In general, we normalize $SU(N)$ generators to
$\tr[t^a(t^b)^\dagger] = \half\delta^{ab}$.)
We can identify the left-handed
electron $e^-_L$, the left-handed positron $e^+_L$, and the photon $A_\mu$ as
\beq
       e^-_L = \s g^{-1}  \ ,   \qquad  e^+_L = \s g^{+1}  \ , \qquad
                      A_\mu = A_\mu^3 \ ,
\eeq{findQED}
where the superscript denotes the matrix from \leqn{threelambdas} which
would be used in computing the Chan-Paton factor.
The three generators form a closed operator algebra, and in fact the
tree amplitudes
of $N=4$ super-Yang-Mills theory which have only these states on external
lines also involve only these states
on internal lines. In string theory,
we can reduce the massless sector to this set of states by an appropriate
orbifold projection \cite{KSO}.  (For example, in a $U(2)$ theory, mod out
by $Z_2\times Z_3$, where $Z_2$ is the  center of $SU(2)$ and the internal
indices $i$ are assigned the $Z_3$ phases $1,\zeta,\zeta,\zeta$, with
$\zeta = e^{2\pi i/3}$).  This gives an explicit prescription
for computing tree-level string corrections to QED amplitudes.
 The electric charge
of the electron  is given by
\beq
 e =  g\ ,
\eeq{evalue}
as one can determine from the commutator $[t^+,t^3]$.  To compute loop
corrections, we should properly extend this theory to a full modular-invariant
string construction.  Instead, for simplicity, we will use the content of
the original $N=4$ supersymmetric theory to compute the loop diagram studied
in Section 6.

Most of our analysis will be carried out at the tree level
in string theory.  A tree-level amplitude
for a particular process actually
depends only on whether that process involves
open- or closed-string states and is otherwise independent of
 which weak coupling string theory
it belongs to.  Beyond this, it
 depends only on the correlation
function of the vertex operators associated with the external particles
for that process and is independent of the remainder of the string spectrum.
If the tree amplitude for a process involves four particles from an
$N=4$ supersymmetric string theory, the amplitude is identical
 whether the full theory has
$N=4$  supersymmetry or is nonsupersymmetric.  This identity is explicit when a
nonsupersymmetric model is constructed as an orbifold of a supersymmetric
theory and, in that situation, is a special case of the `inheritance'
property of orbifolds.  This identity is also familiar in field theory,
  where tree-level
scattering amplitudes in QCD are computed by recognizing that they are
identical to amplitudes in a supersymmetric generalization of QCD \cite{PT}.
Thus, the string corrections to tree-level
Standard Model amplitudes that we will
compute in this paper are actually valid for any situation in which the
quarks and leptons come from the untwisted sector of an open string
orbifold.

Our tree amplitudes are model-independent in another way.  An alternative
string construction of the ADD scenario would be to consider Type IIA string
theory with 5 dimensions large and one dimension small.
Then the ADD scenario would arise if the Standard Model particles
were bound to a D4-brane wrapped around the small dimension.  Similarly, one
could consider $n$ large and $(6-n)$ small dimensions, with a
D$(9-n)$-brane wrapped around the small dimensions.  If the small dimensions
are smaller than 1/TeV, all external states would necessarily
carry zero momentum in these directions.  Then actually the tree
amplitudes derived in this paper would apply for any value of $n$.
We should also note that while we assume the toroidal compactification
of the extra dimensions here, we expect the results for
scattering of open strings on the D brane in Sections 3 and 4 to remain
valid for models with a warp factor in the bulk \cite{RS}, provided that
the bulk curvature is sufficiently small near the brane.

\section{Stringy corrections to $\ee\to \gamma\gamma$ and Bhabha scattering}

In this section, we will use our toy model to compute the effects
of TeV scale strings on Bhabha scattering and $\gamma\gamma$ production in
$\ee$ collisions.  We will compute the leading-order scattering amplitudes
in string theory, using the external states described in the previous section.

Tree amplitudes of open-string theory are given as sums of ordered amplitudes
multiplied by group theory Chan-Paton factors \cite{bigbook}.
We consider amplitudes with all momenta directed inward.
Let the
ordered amplitude with external states $(1,2,3,4)$ be denoted $g^2A(1,2,3,4)$.
Then the full scattering amplitude $\A(1,2,3,4)$ is given by
\beqa
  \A(1,2,3,4) &= &  g^2 A(1,2,3,4) \cdot
             \tr[ t^1 t^2 t^3 t^4 + t^4 t^3 t^2 t^1 ]   \CR
              & & + \  g^2 A(1,3,2,4) \cdot
             \tr[ t^1 t^3 t^2 t^4 + t^4 t^2 t^3 t^1 ]  \CR
              & &  +\   g^2 A(1,2,4,3) \cdot
             \tr[ t^1 t^2 t^4 t^3 + t^3 t^4 t^2 t^1 ] \ .
\eeqa{Aformula}
To compute QED amplitudes with fixed external states,
we would substitute for each $t^i$ the appropriate matrix from
\leqn{threelambdas}
(or, for outgoing states, the Hermitian conjugate matrix).

\begin{figure}
\centerline{\epsfysize=3.000truein \epsfbox{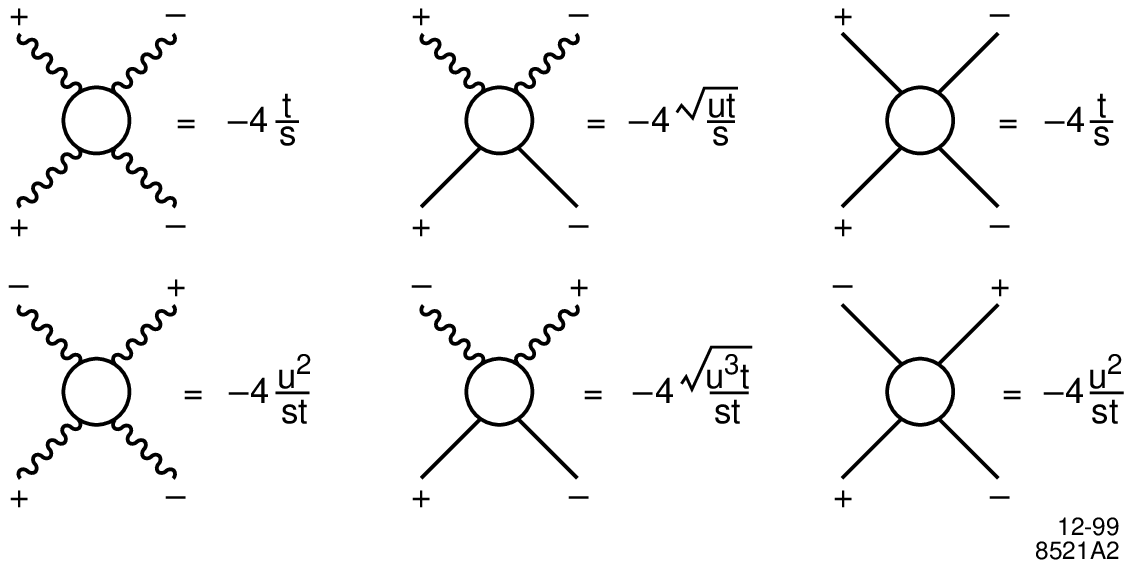}}
\vskip 0.0 cm
 \caption{Nonzero 4-point ordered tree amplitudes of Yang-Mills theory.
 Wavy lines represent gauge bosons; straight lines represent fermions.
 The sign for each line is the helicity, directed inward.}
\label{fig:amps}
\end{figure}

The field theory tree amplitudes of Yang-Mills theory
 can be cast into the same form~\cite{PT},
 and it is useful to consider that case first.  Only a subset of
the possible 4-point ordered amplitudes are nonzero; those amplitudes are
given in Figure~\ref{fig:amps}.  In this figure, a wavy external line
denotes a gauge boson, and a straight external line denotes a fermion.
The sign denotes the helicity (for states directed inward).  The diagrams
are presented with the $s$-channel vertical and the $t$-channel horizontal.
Actually, the four amplitudes involving fermions can be derived from the
two with only gauge bosons by the use of $N=1$ supersymmetry Ward identities,
and these identities also imply the vanishing of the ordered amplitudes for
helicity combinations not shown in the figure.  The two 4-gauge boson
amplitudes are related by $N=2$ supersymmetry.  This is an example of the
model-independence discussed at the end of the previous section.

%%%%%%%%%%%%%%%%%%%%%%%%%%%%%%%%%%%%%%%%%%%%%%%%%%%%%%%%%%%%%%%%%%%%%%
%%%%%%%%%%%%%%%%%%%%%%%%%%%%%%%%%%%%%%%%%%%%%%%%%%%%%%%%%%%%%%%%%%%%%%

It is straightforward to check that these formulae give the familiar
QED tree amplitudes.  For example, for $e^-_L e^+_R$ elastic scattering,
only the first line of \leqn{Aformula} has a nonzero Chan-Paton factor
and we find
\beq
    \A(e^-_Le^+_R \to e^-_L e^+_R) =  - 2e^2 {u^2\over st} =
                  2e^2 u\left({1\over s} + {1\over t}\right) \ ,
\eeq{firstBhabha}
with $g = e$.
For $\ee$ annihilation to $\gamma\gamma$, all three terms contribute and
we find, for example,
\beq
   \A(e^-_Le^+_R \to \gamma_L\gamma_R) =
-e^2 \sqrt{{u\over t}}\left[{u\over s} + {t\over s}  - 1\right] =
 2e^2 \sqrt{{u\over t}} \ .
\eeq{firstgammagamma}

The generalization of the formulae in Figure~\ref{fig:amps} to string
states on a D-brane is known to be quite simple  \cite{Myers,Klebanov} :
 All of the amplitudes shown in the figure are
multiplied by the common factor
\beq
   \S(s,t) = {\G(1-\ap s) \G(1-\ap t) \over
\G(1-\ap s-\ap t)}  \ .
\eeq{Sfactor}
This factor is essentially the original Veneziano amplitude \cite{Venez}.
Before we apply this result, it will be useful to sketch its derivation.

\begin{figure}
\centerline{\epsfysize=1.50truein \epsfbox{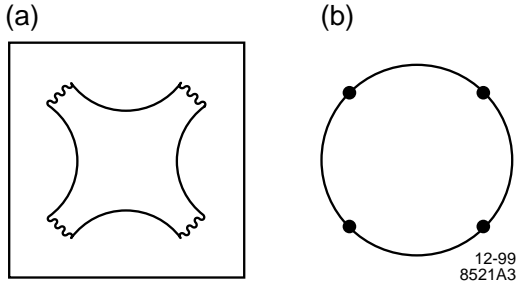}}
\vskip 0.0 cm
 \caption{Schematic illustration and world-sheet diagram of the
scattering process involving four open strings on a D-brane.}
\label{fig:openstrings}
\end{figure}

 In the
model described in Section 2, the electron and photon states are massless
states of open strings ending on the D3-brane.  These states are described
by the quantum theory of fluctuations of an open string in which the string
fields have Neumann boundary conditions in the $\mu =$ 0--3 directions and
Dirichlet boundary conditions in the $\mu =$ 5--10
directions.
The string
world surface has the topology of a disk, as shown in
Figure~\ref{fig:openstrings}(a).  The scattering amplitudes are evaluated
by mapping this surface onto a circle in the complex plane, as in
Figure~\ref{fig:openstrings}(b), and then into the upper half plane.
External open string states are represented by operators,
called `vertex operators', placed on the boundary, and group theory
matrices $t^a$, the Chan-Paton factors.   When the boundary
is mapped to the real line, the vertex
operators appear in a given order 1,2,3,4, and their correlation function
gives the ordered amplitude $A(1,2,3,4)$ which appears in \leqn{Aformula}.
By summing over all orderings, one builds up the complete formula for
$\A(1,2,3,4)$.

%%%%%%%%%%%%%%%%%%%%%%%%%%%%%%%%%%%%%%%%%%%%%%%%%%%%%%%%%%%%%%%%%%%%%%
%%%%%%%%%%%%%%%%%%%%%%%%%%%%%%%%%%%%%%%%%%%%%%%%%%%%%%%%%%%%%%%%%%%%%%

The explicit formula for the 4-point ordered amplitude is  \cite{bigbook,CFT}
\beq
A(1,2,3,4) =   {1\over \ap^2} X^2
\int_{0}^{1} dx \left<\prod_{i=1}^4 {\cal V}_{q_i} (x_i,k_i) \right>\ ,
\eeq{general}
where $\V_{q_i}(x_i)$ is the vertex operator of the state $i$.  The operators
are placed on the real axis at $x_i = 0, x, 1, X$, with $X$ to be fixed and
sent to $\infty$. The index $q_i$ denotes the superconformal charge, which
for the disk amplitude is constrained by $\sum_i q_i = -2$.

A good way to account for the boundary conditions on the real line is
to perform the `doubling trick', which represents left-moving fields on
the world-sheet by fields in the upper half plane and right-moving fields
by their continuation to the lower half plane. Explicitly, let us
split the worldsheet boson field into its holomorphic and antiholomorphic
parts:
\beq
X^\mu(z,\bar{z}) = X^\mu(z) + \bar{X}^\mu (\bar{z}).
\eeq{split}
The boundary conditions imposed on $X(z)$ and the worldsheet fermion
field $\psi(z)$ on the real line are then
\beq
X^\mu(z) = \pm \bar{X}^\mu (\bar{z}), \,\,\,\,
\psi^\mu(z) = \pm \bar{\psi}^\mu (\bar{z}),
\eeq{bcond}
where the plus sign corresponds to $\mu = 0-3$ (Neumann boundary conditions),
and the minus sign to $\mu=5-10$ (Dirichlet boundary conditions.) The
fields $X(z)$ and $\psi(z)$ are originally defined only
on the upper half-plane, $\cal{C}^+$. We extend the definitions of these
fields to the full plane by identifying
\beq
X^\mu(z) = \pm \bar{X}^\mu(z) \ , \qquad
\psi^\mu(z) = \pm \bar{\psi}^\mu (z)\ ,
\qquad  z \in {\cal C}^-,
\eeq{doubling}
where the plus and minus signs again correspond to the Neumann and
Dirichlet boundary conditions. With these definitions, the correlation
functions of these fields are given by
\beqa
\left< X^\mu (w) X^\nu (z) \right> &=&  -{\ap \over 2} g^{\mu\nu}
\ln |w-z|, \CR
\left< \psi^\mu (w) \psi^\nu (z) \right> &=& g^{\mu\nu} (w-z)^{-1},
\eeqa{corr1}
for any $\mu$ and $\nu$.

The open string vertex operators are  built from the
worldsheet boson and fermion fields $X^\mu$ and $\psi^\mu$,
the spin field $\Theta_\alpha$, and the superconformal ghost field $\phi$.
We work in the space-time metric $(-,+,\ldots,+)$, and define
the conventional Mandelstam variables by $s = - 2k_1 \cdot k_2$,
$t=-2k_1 \cdot k_4$, and $u=-2k_1 \cdot k_3$. Then,
for photons, the vertex operators with $q = -1$ and $q=0$ take the form
\beqa
{\cal V}_{-1}^\mu(x,k) &=&  (2 {\ap})^{1/2} e^{-\phi} \psi^\mu
e^{i 2 k\cdot X} (x), \CR
{\cal V}_{0}^\mu(x,k) &=& 2 (i \partial X^\mu +  \ap k \cdot \psi \psi^\mu)
e^{i 2 k\cdot X} (x)\ .
\eeqa{photons}
These expressions are referred to, respectively, as the `$-1$ picture' and
the `0 picture'. The factor of 2 in the exponentials compensates for the
replacement of the full $X^\mu(z,\bar{z})$ by its holomorphic part
in \leqn{split}.
For fermions, the vertex operator with with $q=-1/2$ (`$-1/2$ picture') is
\beq
{\cal V}_{-1/2}^\alpha (x,k) = 2^{1/2} {\ap}^{3/4} e^{-\phi/2}
\Theta^\alpha e^{i 2 k\cdot X} (x) .
\eeq{photinos}
Note that for open strings, the momenta and polarization tensors are
required to be parallel to the D-brane, so all the fields that appear in
the vertex operators \leqn{photons} and \leqn{photinos} have Neumann
boundary conditions. It is then not surprising that the result
\leqn{Sfactor} is identical to the corresponding result in type I
string theory.

The correlators required for the calculation are given by \leqn{corr1}
and
\beqa
\left< e^{-\phi(w)} e^{-\phi(z)} \right> &=& (w-z)^{-1}, \CR
\left< \Theta_\alpha (w) \Theta_\beta (z) \right> &=& C_{\alpha\beta}
(w-z)^{-5/4},
\eeqa{corr2}
where $C_{\alpha\beta}$ is the charge conjugation matrix.
Explicitly evaluating
the expressions \leqn{general} with these vertex operators and
correlators, one finds
the expressions in Figure~\ref{fig:amps} multiplied by the form factor
\leqn{Sfactor}, as promised.

\begin{figure}
\centerline{\epsfysize=1.50truein \epsfbox{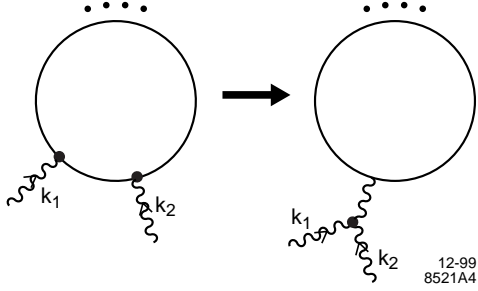}}
\vskip 0.0 cm
 \caption{Factorization of the open-string amplitude to produce a pole
in $(k_1 + k_2)^2$ and a three-gluon vertex.}
\label{fig:factorize}
\end{figure}

A check on the normalization of the 0 picture operator is given by the
operator product relation
\beqa
 \epsilon_2 \cdot{\cal V}_{-1}(x,k_2) \,\,
\epsilon_1 \cdot {\cal V}_0(0,k_1) &\sim& \CR\CR
 & & \hskip -1.85in  - \ap  x^{2 k_1\cdot k_2 \ap-1 }
       \left\{ \epsilon_1 \cdot \epsilon_2 (k_1 - k_2)_\mu +
              2 \epsilon_1\cdot k_2 \epsilon_{2\mu}
              - 2 \epsilon_2\cdot k_1 \epsilon_{1\mu} \right\}
            {\cal V}_{-1}^\mu (0,k_1+k_2) + \Delta  \ ,
\eeqa{VVOPE}
where $\Delta$ is a total derivative in $x$.  A similar  relation holds
for $\V_0(x,k_2)  \V_0(0,k_1) $.  When inserted into \leqn{Aformula}, these
relations give the correct factorization to a pole in $(k_1+k_2)^2$ and the
three-gluon vertex, as shown in Figure~\ref{fig:factorize}.  The
relative normalization of $\V_0$ and $\V_{-1}$ is given by the picture-changing
relation \cite{bigbook}.  Then comparison of the four-point amplitudes
to those of Yang-Mills theory gives the normalization of \leqn{general}.

%%%%%%%%%%%%%%%%%%%%%%%%%%%%%%%%%%%%%%%%%%%%%%%%%%%%%%%%%%%%%%%%%%%%%%
%%%%%%%%%%%%%%%%%%%%%%%%%%%%%%%%%%%%%%%%%%%%%%%%%%%%%%%%%%%%%%%%%%%%%%

To compare string amplitudes to Standard Model amplitudes, we are
typically interested in the limit
in which $s$, $t$, $u$ are much less than the string scale $M_S = \ap^{-1/2}$.
In this limit,
\beq
\S(s,t) = \left(1 - {\pi^2\over 6}  {st\over M_S^4} + \cdots \right) \ .
\eeq{primexpand}
It is interesting that, in the toy model, the leading corrections are
proportional to $M_S^{-4}$, corresponding to an operator of dimension~8.
This is a consequence of the fact that the first higher-dimension operator
with $N=4$ supersymmetry appears at dimension 8 \cite{Tseytlin}.  It is
likely that in more general string models in which quarks and leptons
appear from twisted sectors of the orbifold, the first string corrections
would be proportional to $M_S^{-2}$.

Now we can apply the form factor \leqn{Sfactor} to representative QED
processes.  For Bhabha scattering, only the first Chan-Paton factor is
nonzero, and so we find
\beqa
    \A(e^-_Le^+_R \to e^-_L e^+_R) &=&  - 2e^2 {u^2\over st} \S(s,t) \ , \CR
    \A(e^-_Le^+_R \to e^-_R e^+_L) &=&  - 2e^2 {t\over s} \S(s,t) \ , \CR
    \A(e^-_Le^+_L \to e^-_L e^+_L) &=&  - 2e^2 {s\over t} \S(s,t) \ ,
\eeqa{stringBh}
and the same results for the parity-reflected processes.
In  general,  all helicity amplitudes for Bhabha scattering
are given by their field theory expressions multiplied by $\S(s,t)$.
This form factor has SR poles in the $s$- and $t$-channels.  A $u$-channel
pole cannot appear, because the open string contains no states with
electric charge $\pm 2$.

For $\ee\to \gamma\gamma$, the result is more complex.  The string form factor
appears in all three possible channels, and we find
\beq
   \A(e^-_Le^+_R \to \gamma_L\gamma_R) =
e^2 \sqrt{{u\over t}}\left[{u\over s}\S(s,t) + {t\over s}\S(s,u)
                                          - \S(t,u) \right]  \ .
\eeq{eeggS}
The other nonzero helicity amplitudes are derived from this one by parity
reflection and crossing.  In particular, the amplitude for production of
$\gamma_R\gamma_R$ remains zero.  The amplitude \leqn{eeggS}
contains massive SR
poles in all three channels.

\section{String phenomenology at colliders}

The expressions for stringy corrections that we have derived
allow one to search for signals of string theory in collider
experiments.  In this section, we will discuss these explicit signatures of
string theory.  We begin by considering effects visible as contact interactions
well below the string scale.  We will then   discuss direct observation of
the string Regge excitations.

\subsection{Contact interactions}

Both two-photon production and Bhabha scattering have been studied at
LEP 2 at the highest available energies.  We consider first the case of
two-photon production.  Deviations from the Standard Model cross section
have been analyzed by the LEP experiments in terms of Drell's
parametrization \cite{Drell}
\beq
 {d \sigma\over d \cos\theta} =
         {d \sigma\over d \cos\theta}\biggr|_{SM} \cdot
   \left(1 \pm  {2ut \over \Lambda_\pm^4}\right) \ .
\eeq{Sidsform}

For the case of $\ee\to \gamma\gamma$, it is actually a general result that
the first correction due to a higher-dimension operator comes from a
unique dimension-8 operator.  This operator is
proportional to the cross term in  $T^{\mu\nu} T_{\mu\nu}$, where
$T^{\mu\nu}$ is the
energy-momentum tensor of QED.  Thus, Drell's  parametrization \leqn{Sidsform}
should apply to any model of new physics at short distances.

To compare our string theory results to this expression,
 insert \leqn{primexpand} into \leqn{eeggS};
 this gives
\beq
   \A(e^-_Le^+_R \to \gamma_L\gamma_R) =
-2e^2 \sqrt{{u\over t}}\left[ 1 + {\pi^2\over 12 } {ut\over M_S^4}  +
                 \cdots \right] \ .
\eeq{eeggSex}
Squaring this expression, and noting that the correction is invariant to
crossing
$t \leftrightarrow u$, we can identify
\beq
      \Lambda_+ = \left( 12/\pi^2 \right)^{1/4} M_S   \ .
\eeq{equateLam}
The OPAL collaboration \cite{OPALgg}
has reported a limit $\Lambda_+ > 304$ GeV from
measurements at 183 and 189 GeV in the center of mass.  The
ALEPH, DELPHI, and L3 collaborations have reported similar
constraints \cite{ALEPHgg,DELPHIgg,L3ee}.
 The OPAL result corresponds to a limit
\beq
        M_S >  290  \ \mbox{GeV} \ , \  \mbox{95\% conf.}
\eeq{MSlimgg}
If we use the first line of
\leqn{MMS} to convert this to a limit on the fundamental quantum gravity
scale, we find $M > 870$ GeV.

The comparison of string predictions to the data on Bhabha scattering
brings in two new considerations.  The first of these is that Bhabha
scattering at energies above the $Z^0$ resonance includes $Z^0$ exchange
as an important contribution, while the $Z^0$ was not a part of our
string QED.  To find a prescription for including both $\gamma$ and
$Z^0$ exchange, we recall that  all QED Bhabha scattering amplitudes
are multiplied by the common form factor $\S(s,t)$. Thus, we suggest
comparing the data on Bhabha scattering to the
simple formula
\beq
   {d \sigma\over d \cos\theta} (e^-e^+\to e^-e^+)  =
         {d \sigma\over d \cos\theta}\biggr|_{SM} \cdot
\left| \S(s,t) \right|^2 \ .
\eeq{SMprocesses}
This is essentially the assumption that the SR excitations of the
photon and the $Z^0$ have the same spectrum, up to contributions of size
$\mz^2$ that we can ignore in computing their masses, and that the SR
excitations of the $Z^0$ have the same polarization asymmetry as the $Z^0$
in their coupling to electrons.

The second complication  for
 Bhabha scattering is that,
 unlike the case of $\ee\to \gamma\gamma$, there are
many possible forms for the higher-dimension corrections to the Standard
Model result.  Already at dimension 6 there are three possible
helicity-conserving
operators, of which two are also
parity-conserving.  At dimension 8 there are 4
parity-conserving operators.  Various combinations of these operators
have been proposed as the basis for fits to Bhabha scattering data.
It would be useful to review the most important models proposed
previously and to compare them to \leqn{SMprocesses}.

For many years, Bhabha scattering has been of interest as the most sensitive
probe of lepton substructure.  The form proposed for deviations from
the Standard Model prediction was the most general combination
of helicity-conserving dimension-6 operators \cite{ELP}
\beq
   \delta\L = {4\pi\over 2 \Lambda^2} \left[
     \eta_{LL} \bar e_L\gamma^\mu e_L \bar e_L \gamma_\mu e_L
   +  \eta_{RR} \bar e_R\gamma^\mu e_R \bar e_R \gamma_\mu e_R
   + 2  \eta_{RL} \bar e_R\gamma^\mu e_R \bar e_L \gamma_\mu e_L\right]  \ ,
\eeq{ELPform}
where the $\eta_a$ are $\pm 1$ or 0 and the mass scale $\Lambda$ is taken
to be the scale of compositeness.

With the recent interest in large
extra dimensions and low-scale quantum gravity, Bhabha scattering has been
reconsidered as a place to look for the effects of virtual KK graviton
exchange.  As we have remarked in the introduction, the effect of KK
exchange is not reliably computable in low-energy effective field theory.
Typically, this effect is modeled by introducting an appropriate contact
interaction with an adjustible coefficient \cite{Jim,Hewett,HLZ}.  In
this paper we will follow Hewett's convention
 by representing the effective Lagrangian
for KK exchange as \cite{Hewett}:
\beq
      \delta \L =   i {4\lambda\over M_H^4} T^{\mu\nu} T_{\mu\nu} \ ,
\eeq{TTint}
where $\lambda = \pm1$ and $T^{\mu\nu}$ is the full energy-momentum tensor
of the model.  Hewett writes the scale in this Lagrangian as $M_S$; we
use the notation $M_H$ to distinguish this mass scale from the string scale
\cite{conventions}.

It should be noted that the expressions \leqn{ELPform} and \leqn{TTint}
do not contain any powers of a small coupling constant. When these expressions
are added to the Standard Model formulae, the higher-dimension operators
compete with amplitudes that are of order $g^2$.  This allows one to obtain
very stringent bounds on the coefficient of the new operators.  Bounds on
the $\Lambda$ parameters, for example, are typically a factor of 20 higher
than the center-of-mass energy of the $\ee$ collisions being analyzed.
The physical meaning of these bounds, however, depends on
the relation between the coefficients in  \leqn{ELPform} and \leqn{TTint}
and the predictions of the underlying fundamental theory.  In Section 6,
we will derive \leqn{TTint} from our toy string model and show that the
coefficient is of order
\beq
               {1\over M_H^{4}} \sim  {g^4 \over M_S^4}  \  .
\eeq{MHrel}
Thus, \leqn{TTint} is parametrically suppressed with respect to
the effects of SR exchange.
This conclusion is generic when quantum gravity is represented by a
weakly-coupled string theory, though perhaps in other models of quantum
gravity \leqn{TTint} might be the dominant effect.

With this in mind, we will compare the models
discussed above to an illustrative data set for Bhabha scattering at
LEP 2.  A complete analysis of
the LEP 2 data is beyond the scope of this paper.  For reference,
we have listed the
various expressions for the Bhabha scattering cross sections in these
models in Appendix A.

%%%%%%%%%%%%%%%%%%%%%%%%%%%%%%%%%%%%%%%%%%%%%%%%%%%%%%%%%%%%%%%%%%%%%%
\begin{figure}
\centerline{\epsfysize=4.00truein \epsfbox{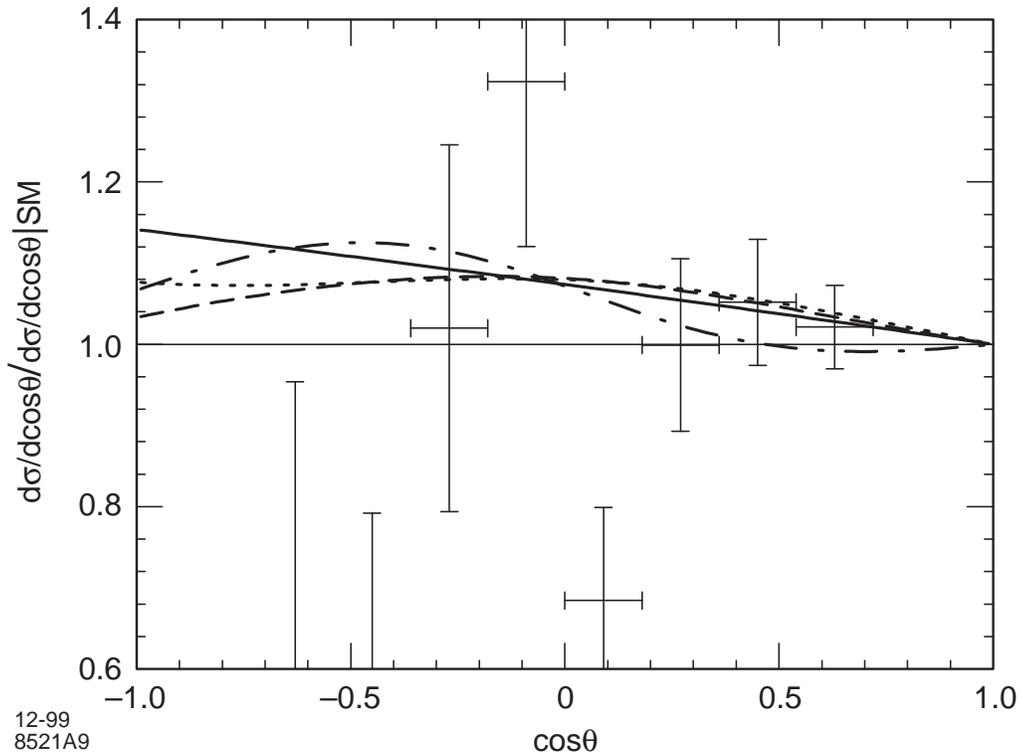}}
\vskip 0.0 cm
\caption{Comparison of data on Bhabha scattering at 183 GeV with models of
corrections to the Standard Model from higher-dimension operators.  The
plot shows the fractional deviation from the Standard Model,
$(d \sigma/d\cos\theta / d \sigma/d\cos\theta/|_\SM -1)$ versus
 $\cos\theta$.  The four curves represent:  solid, string model with
$M_S = 410$ GeV; dotted, KK exchange with $M_H = 830$ GeV;
dashed, VV contact interactions with $\Lambda = 8800$ GeV;
dot-dash, AA contact interactions with  $\Lambda = 6700$ GeV.}
\label{fig:Lthree}
\end{figure}
%%%%%%%%%%%%%%%%%%%%%%%%%%%%%%%%%%%%%%%%%%%%%%%%%%%%%%%%%%%%%%%%%%%%%%

The four LEP experiments have all announced preliminary results on
 the Bhabha scattering cross
section at high energies \cite{L3ee,ALEPHee,OPALee,DELPHIee,Bourilkov}
and have used the results to put limits on
4-fermion contact interactions.  In particular,
the L3 experiment has published
their data at 183 GeV in a form convenient for our analysis.
In Figure~\ref{fig:Lthree}, we compare this data to the formula
\leqn{SMprocesses} and to the analogous formulae derived from
\leqn{ELPform} and \leqn{TTint}.  The curves shown are the
95\% confidence exclusion limits for the various models considered:
for SR exchange,
$M_S > 410$ GeV, for KK exchange with $\lambda = +1$,
$M_H > 830$ GeV,  for compositeness with VV contact interactions
($\eta_{LL} = \eta_{RR} = \eta_{RL} = -1$)
$\Lambda >  8800$ GeV, for compositeness with AA contact
interactions,
($\eta_{LL} = \eta_{RR} = -\eta_{RL} = +1$),
$\Lambda >  6700$ GeV.   In a weakly-coupled string theory, the dominant
effect would come from $M_S$.  Using the relation \leqn{MMS},
 the exclusion limit on $M_S$ derived from this data would correspond to a
limit on the quantum gravity scale of $M > 1.2$ TeV.

\begin{figure}
\centerline{\epsfysize=4.00truein \epsfbox{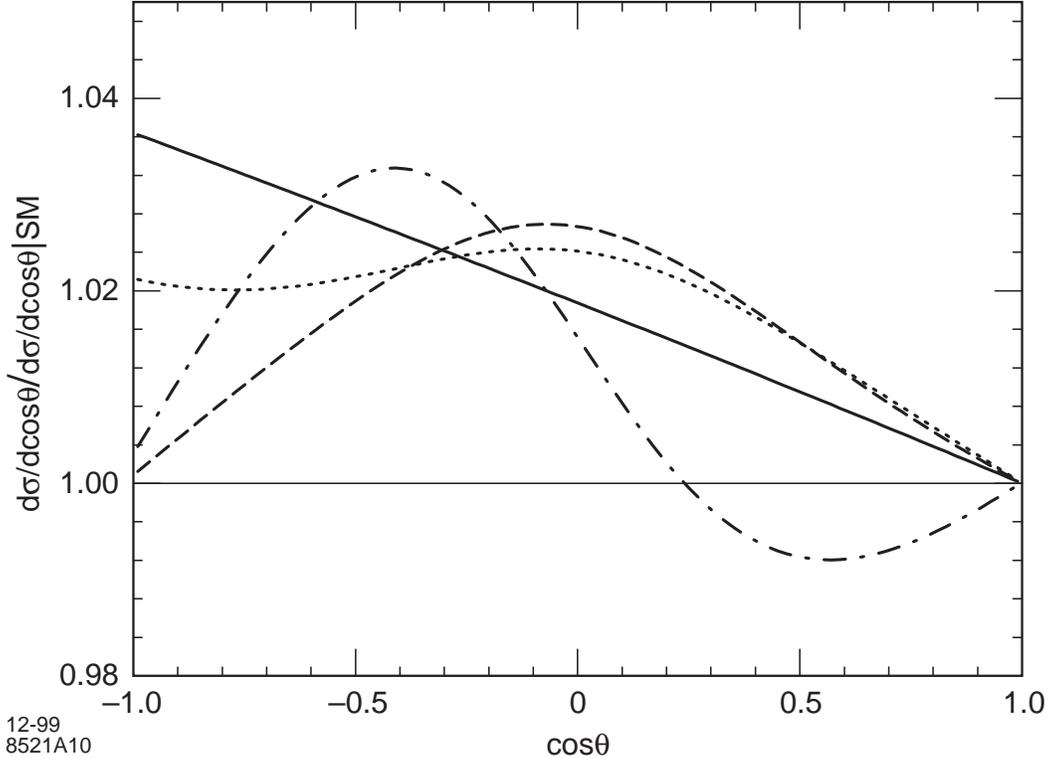}}
\vskip 0.0 cm
\caption{Comparison of deviations from the Standard Model prediction for
Bhabha scattering  at 1 TeV due to
corrections from higher-dimension operators.
  The four curves represent:  solid, string model with
$M_S = 3.1$ TeV; dotted, KK exchange with $M_H = 6.2$ TeV;
dashed, VV contact interactions with $\Lambda = 88$ TeV;
dot-dash, AA contact interactions with  $\Lambda = 62$ TeV.}
\label{fig:NLC}
\end{figure}

A similar analysis can be used to estimate the sensitivity of experiments at
future, higher-energy $\ee$ colliders.  As a guide, consider a linear
$\ee$ collider running at a center of mass energy of 1 TeV.  With a
100~fb$^{-1}$ data sample, the measurement of Bhabha scattering should be
systematics limited.  We consider a set of 8 measurements of the
differential cross sections corresponding to the bin centers in
Figure~\ref{fig:Lthree} and assume that each measurement is made to 3\%
accuracy and agrees with the Standard Model expectation.
Then the 95\% confidence exclusion limits for the four models
just considered are: for SR exchange,
$M_S > 3.1$ TeV, for KK exchange with $\lambda = +1$,
$M_H > 6.2$ TeV,  for compositeness with VV contact interactions
$\Lambda >  88$ TeV, for compositeness with AA contact
interactions,
$\Lambda >  62$ TeV.  The corresponding deviations from the Standard
Model expectation are graphed as a function of $\cos\theta$ in
Figure~\ref{fig:NLC}.
Using \leqn{MMS}, the limit on $M_S$ would translate to
a limit $M > 9.3$ TeV on the quantum gravity scale.

A remarkable feature of Figure~\ref{fig:NLC} is that the four curves shown
have very different shapes.  If a deviation from the Standard Model is
seen, then with higher statistics or higher energy it should be possible to
determine which of these theories, if any, gives the correct description.

%%%%%%%%%%%%%%%%%%%%%%%%%%%%%%%%%%%%%%%%%%%%%%%%%%%%%%%%%%%%%%%%%%%%%%
%%%%%%%%%%%%%%%%%%%%%%%%%%%%%%%%%%%%%%%%%%%%%%%%%%%%%%%%%%%%%%%%%%%%%%

\subsection{Resonances}

Though theories based on contact interactions are limited to the first
deviations from the Standard Model, our string theory formulae are valid
at higher energies, and we can examine their characteristic features there.
The most obvious property apparent in \leqn{Sfactor} is the presence of a
sequence of $s$-channel poles at masses $M_n = \sqrt{n} M_S$,
 for $n = 1,2, \ldots$.  It is interesting to explore the properties of the
first resonances in some detail.

The stringy form factor $\S(s,t)$ has its first pole at $s= M_S^2$.  Near
this point, it has the form
\beq
          \S(s,t) \sim  {t\over s - M_S^2}  \ .
\eeq{Spole}
We can use \leqn{Spole} to find the first resonance in string QED tree
amplitudes.  The pole contributions are
\beqa
 \A(e^-_Le^+_R \to e^-_Le^+_R) &=& - 2e^2 {u^2\over s^2} {s \over s - M_S^2},
                                                   \hskip1.6cm
 \A(e^-_Le^+_R \to e^-_Re^+_L) = - 2e^2 {t^2\over s^2} {s \over s - M_S^2},
                                                   \CR
 \A(e^-_Le^+_R \to \gamma_L \gamma_R) &=&
                       - 2e^2 {u \sqrt{ut}\over s^2} {s \over s - M_S^2},
                                                   \hskip1cm
 \A(\gamma_L \gamma_R \to \gamma_L \gamma_R) =
                       - 2e^2 {u^2\over s^2} {s \over s - M_S^2},
                                                   \CR
\A(\gamma_R \gamma_R \to \gamma_R\gamma_R) &=&
                       - 2e^2 {s \over s - M_S^2},
                                                   \hskip2cm
\A(e^-_R e^+_R \to e^-_R e^+_R) =
                       - 2e^2 {s \over s - M_S^2},    \CR
\eeqa{allpoles}
with equal results for the parity-reflected and time-reversed processes, and
zero for all other possible reactions.

\begin{figure}
\centerline{\epsfysize=6.00truein \epsfbox{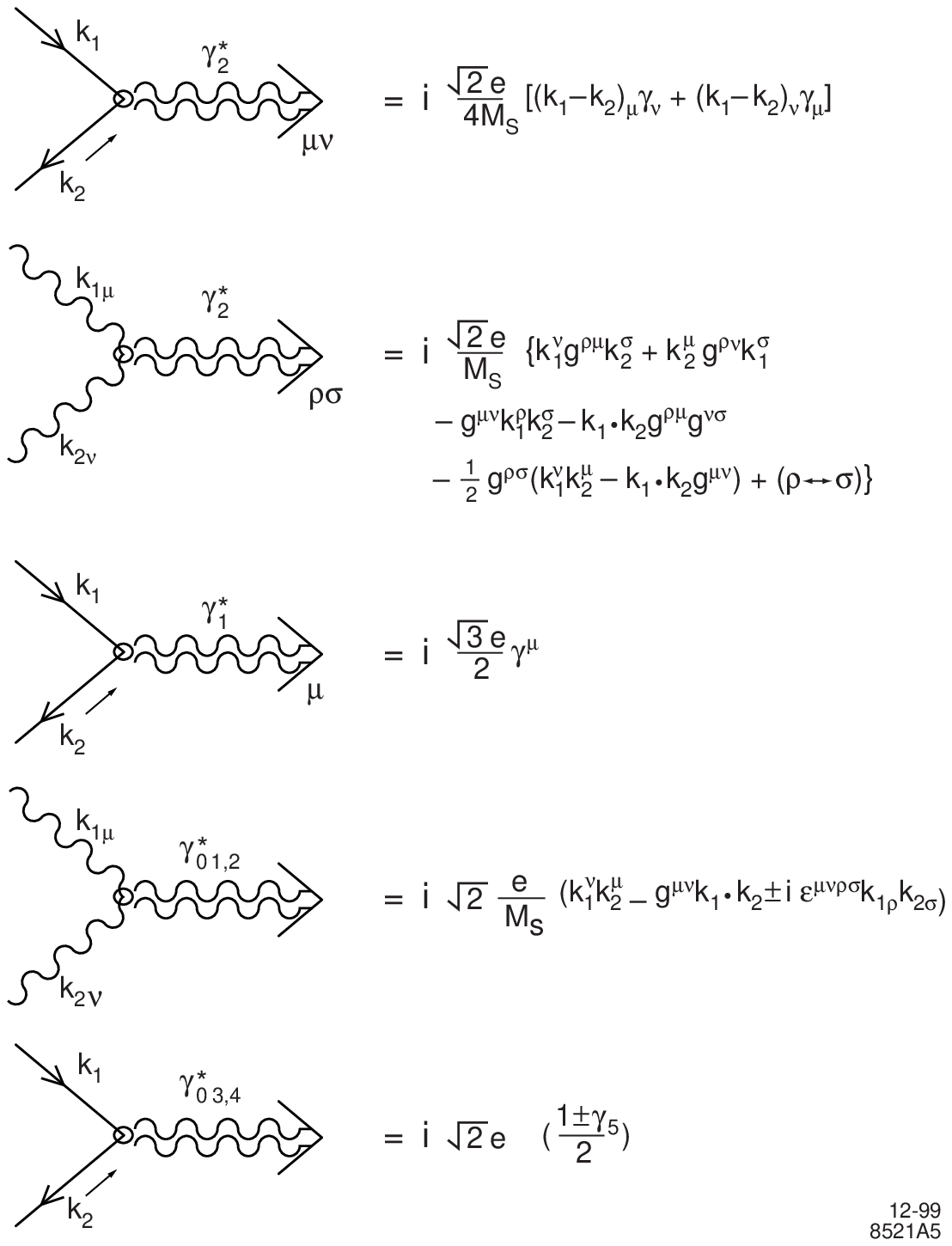}}
\vskip 0.0 cm
\caption{Feynman rules for the coupling of the  SR resonances at $M = M_S$
          in string QED to electron and photon pairs.}
\label{fig:Feynmans}
\end{figure}

The properties of the first SR resonances can then be found by factorizing
these expressions.  They require four spin 0 resonances $\gamma_{0i}$,
$i = 1, \ldots, 4$,
 one spin 1 resonance $\gamma_1^*$ and one spin 2
resonance $\gamma_2^*$.   Four spin zero resonances are
needed because the transition amplitudes between any pair of
$e^-_Re^+_R$,  $e^-_Le^+_L$,
$\gamma_R\gamma_R$ and $\gamma_L\gamma_L$ vanish.  The on-shell
couplings of electron and photon pairs to the resonances are
\beqa
\begin{tabular}{l l}
  $\A(\gamma_R\gamma_R \to \gamma_{01}^*) = \sqrt{2} e M_S,$  & \hskip55pt
  $\A(e^-_L e^+_R \to \gamma_1^*)  = \sqrt{{3\over 2}} e M_S \epsilon_-^\mu,$\CR

  $\A(\gamma_L\gamma_L \to \gamma_{02}^*) = \sqrt{2} e M_S,$  & \hskip55pt
  $\A(e^-_R e^+_L \to \gamma_2^*)  = \sqrt{{1\over 2}} e M_S
  \cdot {1\over \sqrt{2}} [   \epsilon_+^\mu \epsilon_0^\nu +
             \epsilon_+^\nu \epsilon_0^\mu ],$    \CR

  $\A(e^-_R e^+_R \to \gamma_{03}^*)  = \sqrt{2} e M_S,$ & \hskip55pt
  $\A(e^-_L e^+_R \to \gamma_2^*)  = \sqrt{{1\over 2}} e M_S
  \cdot {1\over \sqrt{2}} [   \epsilon_-^\mu \epsilon_0^\nu +
             \epsilon_-^\nu \epsilon_0^\mu ],$ \CR

  $\A(e^-_L e^+_L \to \gamma_{04}^*)  = \sqrt{2} e M_S,$ & \hskip55pt
  $\A(\gamma_L \gamma_R \to \gamma_2^*) = \sqrt{2} e M_S
                                       \epsilon_-^\mu \epsilon_-^\nu,$  \\
  $\A(e^-_R e^+_L \to \gamma_1^*) = \sqrt{{3\over 2}} e M_S
                                                      \epsilon_+^\mu,$ \\
\end{tabular}
\hskip2pt\\
\eeqa{alldecays}

where, when the first particle moves in the $+\hat 3$ direction,
\beq
  \epsilon_+^\mu = {1\over \sqrt{2}} (0,1,i,0)^\mu \ ,\qquad
  \epsilon_-^\mu = {1\over \sqrt{2}} (0,1,-i,0)^\mu \ ,\qquad
  \epsilon_0^\mu =  (0,0,0,1)^\mu \ .
\eeq{epsilondefs}
Feynman rules which give rise to these expressions are listed in
Figure~\ref{fig:Feynmans}.

%%%%%%%%%%%%%%%%%%%%%%%%%%%%%%%%%%%%%%%%%%%%%%%%%%%%%%%%%%%%%%%%%%%%%%
%%%%%%%%%%%%%%%%%%%%%%%%%%%%%%%%%%%%%%%%%%%%%%%%%%%%%%%%%%%%%%%%%%%%%%

From these expressions, we can compute the width of the resonances.
For the scalar SR resonances,
\beq
    \Gamma_{01} = \Gamma_{02} =   {\alpha\over 4}M_S \ , \qquad
    \Gamma_{03} = \Gamma_{04} =   {\alpha\over 2}M_S  \  .
\eeq{scalargams}
For the vector resonance,
\beq
      \Gamma_1 = {\alpha\over 4} M_S \ ,
\eeq{vectorgams}
with equal contributions from decays to $e^-_Re^+_L$ and
$e^-_Le^+_R$.  For the spin 2 resonance,
\beq
   \Gamma_2(e^+e^-) = \Gamma_2(\gamma\gamma) =  {\alpha\over 20} M_S \ ,
         \qquad \Gamma_2 =  {\alpha\over 10} M_S \ ,
\eeq{tensorgams}
again with equal contributions from  $e^-_Re^+_L$ and
$e^-_Le^+_R$.  The production cross sections can be derived from these
formulae using, for example
\beq
       \sigma(e^+e^- \to \gamma^*_J) = 4 \pi^2 (2J+1) {\Gamma(\gamma_J^* \to
               e^+e^-)\over M_S} \delta(s - M_S^2) \ .
\eeq{crosssecgamma}

In $\ee$ collisions, one currently has data available only up to
200 GeV.  In quark-antiquark processes, however, collision energies
up to 1 TeV and above are available in the Tevatron data.  Thus, it is
important to generalize this analysis to $q\bar q$ collisions so that we
can ask whether the SR excitations of the gluon ought to have been seen
at the Tevatron.  We will now present our first attempt at a generalization
of string QED to string QCD.  Though this theory will not be completely
satisfactory, it will at least allow us to estimate the bound on the
string scale from the study of jets at the Tevatron.

Consider, then, a system of four D3-branes with a $U(4)$ gauge symmetry.
Represent the gluons of QCD by the gauge bosons of $SU(3) \subset U(4)$,
that is, by $3\times 3$ Chan-Paton matrices $t^a$.
Represent left-handed quarks and antiquarks of one flavor by the $U(4)$
matrices
\beq
           (t^i)_{pq} = {1\over \sqrt{2}} \delta^i_p \delta^4_q \ , \qquad
          (\bar t^i)_{pq} = {1\over \sqrt{2}} \delta^i_q \delta^4_p \ .
\eeq{quarkmatrices}
Ideally, we would like to make an orbifold projection of the $U(4)$ theory
onto a theory which contained only these quarks and gluons at the massless
level.  Unfortunately, this is not possible,
because the commutator $[t^i, \bar t^j]$ includes
not only  a linear combination of the $t^a$ but also the $U(1)$ generator
\beq
             t_4 = {1\over \sqrt{24}}
 \pmatrix{1 & &&\cr  & 1 & &\cr  & &1&  \cr &&& -3\cr} \ .
\eeq{UoneinUfour}
Thus,
this massless $U(1)$ gauge boson will also appear in quark-quark scattering
amplitudes.

Keeping  this problem in mind, we
 compute the amplitude for $q_L \bar q_R$ scattering
using \leqn{Aformula}.  Only the first line has a nonzero Chan-Paton factor,
which equals
\beqa
 \tr[ t^i \bar t^j t^k \bar t^\ell + \bar t^\ell t^k \bar t^j t^i ]
 &=& {1\over 4} \left\{ \delta^{jk} \delta^{\ell i} +
                             \delta^{k\ell} \delta^{ij} \right\} \CR
 &=& {1\over 2} \left\{ (t^a)_{ji} (t^a)_{\ell k} + {2\over 3}
            \delta^{ji} \delta^{\ell k} \right\} \ .
\eeqa{colorfactor}
In the last line, the first term corresponds to color octet exchange in the
$s$-channel, and the second term to exchange of a $U(1)$ boson corresponding
to the generator \leqn{UoneinUfour}.  To make our estimate, we will drop the
$U(1)$ piece and then factorize the color octet piece of the amplitude
as above.  This gives
\beq
   {\cal A}(q_L^i \bar q_R^j \to q_L^\ell \bar q_R^k) =
           -2g^2 {u^2\over st}  (t^a)_{ji} (t^a)_{\ell k}  \cdot \S(s,t)\ ,
\eeq{qamp}
which implies:
\beqa
  \A(q_L^i \bar q^j_R \to g_1^{*a}) & =& \sqrt{{3\over 2}} g M_S (t^a)_{ji}
                                                      \epsilon_-^\mu \ , \CR
  \A(q^i_L \bar q^j_R \to g_2^{*a}) & =& \sqrt{{1\over 2}} g M_S (t^a)_{ji}
  \cdot {1\over \sqrt{2}} [   \epsilon_-^\mu \epsilon_0^\nu +
             \epsilon_-^\nu \epsilon_0^\mu ] \ ,
\eeqa{qres}
and similarly for $q_R^i \bar q^j_L$.  The result is just what we would
have obtained by replacing $e$ by $g$ and adding an $SU(3)$ color matrix
in  the Feynman rules of Figure~\ref{fig:Feynmans}.
 From these matrix elements, we can
compute the production cross sections from unpolarized $q\bar q$ initial
states:
\beq
   \sigma(q\bar q \to g_1^*) = {4 \pi^2 \alpha_s \over 3} \delta(s-M_S^2)\ ,
             \qquad
   \sigma(q\bar q \to g_2^*) = {4 \pi^2 \alpha_s \over 9} \delta(s-M_S^2)\ ,
\eeq{gncross}
so that
\beq
   \sigma(q\bar q \to g^*) = {16 \pi^2 \alpha_s \over 9} \delta(s-M_S^2)\ .
\eeq{gnncross}

The result  \leqn{gnncross}
can be compared to the cross section for producing the
axigluon \cite{axig} and  coloron \cite{coloron}, hypothetical massive
vector or axial vector  bosons
that couple to $q\bar q$ with the QCD coupling strength.  In either case,
the cross section is
\beq
   \sigma(q\bar q \to V) = {16 \pi^2 \alpha_s \over 9} \delta(s-M_S^2)\ .
\eeq{axigluon}
Then we can use experimental constraints on these objects to place a direct
experimental bound on the string scale.  A recent paper by the
CDF collaboration has searched for
 the presence of a narrow resonance in the two-jet invariant mass
 distribution in $p\bar p$ collisions at the
Tevatron \cite{CDFres}.  The CDF collaboration does not find evidence for such
a resonance and puts a lower limit of 980 GeV (at 95\% conf.) on the axigluon
or coloron mass.  Naively, we should have the same limit on $M_S$.
 Several uncertain factors appear in this comparison, however.
 On the negative side, the events with $g_2^*$  have an angular distribution
which is more peaked toward the beam axis, and so the acceptance for these
events should be lower.  The angular distribution for the $g_1^*$ events
is identical to that from the axigluon or coloron.
On the positive side, we have ignored scalar gluon
resonances and the production of $g_1^*$ and $g_2^*$ by gluons.  Thus, we
might say that the CDF limit constraints the string scale $M_S$ to be
greater than approximately 1 TeV.  If we convert this limit to a limit
on the quantum gravity scale
 using the second line of  \leqn{MMS}, we find that $M > 1.6$ TeV.

%%%%%%%%%%%%%%%%%%%%%%%%%%%%%%%%%%%%%%%%%%%%%%%%%%%%%%%%%%%%%%%%%%%%%%
\begin{figure}
\centerline{\epsfysize=4.000truein \epsfbox{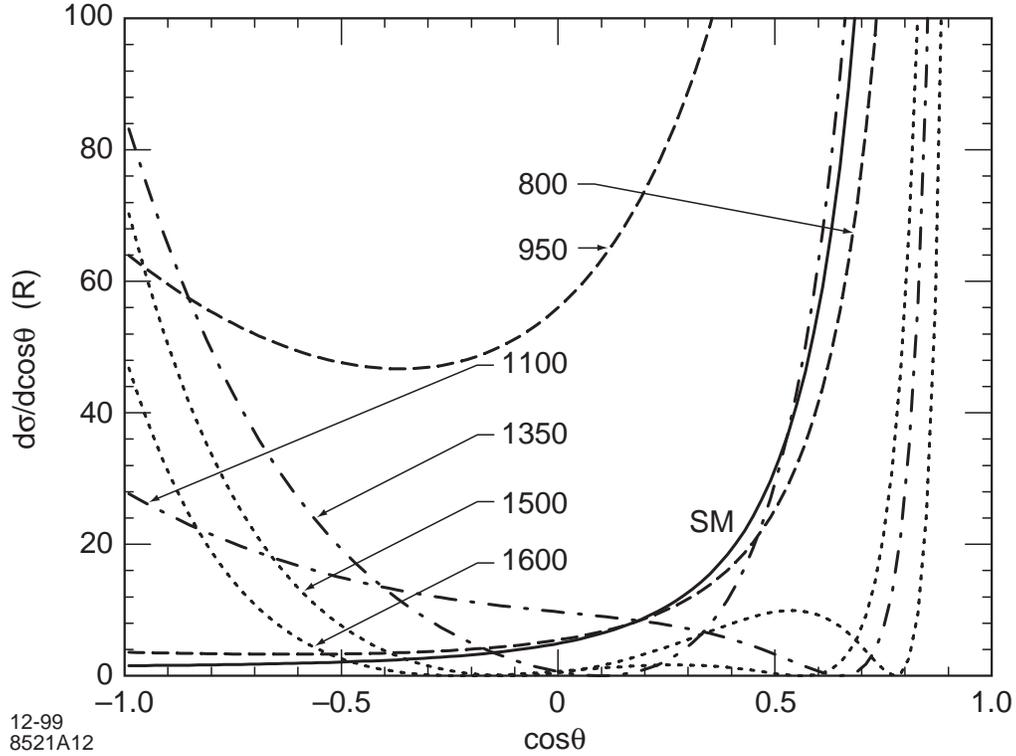}}
\vskip 0.0 cm
\caption{Differential cross section for Bhabha scattering in our string model,
with $M_S = 1$ TeV, at a sequence of center of mass energies that interleave
the first few resonances.  The cross sections are given in units of $R =
4\pi\alpha^2/3s$.
The number next to each curve indicates the
energy.  The various line types show: solid, Standard Model prediction;
dashed, $E_\CM < M_S$;  dot-dash, $M_S < E_\CM < \sqrt{2} M_S$; dotted,
 $\sqrt{2} M_S < E_\CM < \sqrt{3} M_S$.}
\label{fig:highB}
\end{figure}
%%%%%%%%%%%%%%%%%%%%%%%%%%%%%%%%%%%%%%%%%%%%%%%%%%%%%%%%%%%%%%%%%%%%%%

The sensitivity to SR resonances in quark and gluon scattering will increase
dramatically when the LHC begins operation.  The sensitivity of higher-energy
hadron colliders to the axigluon was estimated some time ago
by Bagger, Schmidt, and King
\cite{BSK}.   Scaling their results to the LHC energy, we expect that the
LHC could put a limit of about 5 TeV on the axigluon mass, and a comparable
limit on $M_S$.  Using \leqn{MMS}, this would correspond to a limit
$M > 8$ TeV.   These values are sufficiently high that string resonances
ought to be discovered at the LHC if the low quantum gravity scale is
connected to the mechanism of electroweak symmetry breaking as suggested by
 ADD \cite{Nima1}.

To conclude this section, we discuss
what happens when we probe even higher energies, above the scale of
the first SR resonance.
 When $s > M_n^2$, the
expression \leqn{Sfactor} has a zero at $t = - (s - M_n^2)$.  Thus, above
the first resonance, there is one zero in $\cos\theta$, above the second
resonance, there are two zeros, and so forth.  This leads to
an angular distribution of the sort produced by diffractive scattering.
In Figure~\ref{fig:highB}, we plot the differential cross section for Bhabha
scattering, from \leqn{SMprocesses}, for a sequence of energies that interleave
the SR resonances.

It is well-known from the old string literature that the
differential cross sections at very high energy have the form of a narrow
diffractive peak.  Indeed, using Stirling's formula to evaluate $\S(s,t)$
in the limit $s\to \infty$ and fixed angle, we find \cite{bigbook}
\beq
     \S(s,t) \sim \exp[-\ap s f(\cos\theta)] \ ,
\eeq{Psform}
where $f(\theta)$ is the positive function
\beq
          f(c) = - {1+c\over 2} \log  {1+c\over 2}
                 - {1-c\over 2} \log  {1-c\over 2}\ .
\eeq{fthetaform}
However, at intermediate energies, the large positive
deviation in the backward direction is also an important part of the string
signature.   As $\cos\theta \to -1$,
\beq
  \bigl|\S(s,t) \bigr|^2 \to \left({\pi \ap s \over \sin\pi\ap s}\right)^2\ .
\eeq{backwards}
Thus for increasing $s$
 there is a larger enhancement, but in a narrower region of
backward angles.

\section{Stringy corrections to $\ee\to \gamma G$}

Our toy model includes the process of
 graviton emission in electron-positron annihilation,
 $\ee\to \gamma G$.  This process gives a missing-energy signature
which becomes significant when the center-of-mass energy of the annihilation
approaches the gravitational scale $M$.  The process has been used by the LEP 2
experiments to put constraints on the size of large extra dimensions.
In this section, we study the stringy corrections to this process.

To begin, we recall that the leading contribution to this process at low
energy is model-independent.  The calculation uses only the fact that
a graviton---even a KK excitation---couples to the energy-momentum tensor
of matter \cite{Raman}.  The coupling has the usual 4-dimensional gravitational
strength.  From this, one finds that the polarized differential cross section
 for the process $\ELER\to \gamma G$, for production of a given KK excitation
of mass $m$, is given by \cite{Jim,MPP}
\beqa
{d\sigma\over d\cos\theta}\biggr|_{\mbox{\scriptsize ft}}
  &=& {\pi \alpha G_N \over 1-m^2/s}\Biggl[(1 +
   \cos^2\theta )\left(1 + \bigl({m^2\over s}\bigr)^4\right) \CR
     & &   + \left( { 1 - 3 \cos^2\theta + 4 \cos^4\theta\over
          1- \cos^2\theta}\right){m^2\over s }\left(1 +
           \bigl({m^2\over s}\bigr)^2\right)
     + 6 \cos^2\theta \bigl({m^2\over s}\bigr)^2
                                          \Biggr] \ .
\eeqa{eegGfield}
To obtain the full cross-section for graviton emission at a given
collision energy, we need to sum over all the modes whose emission is
kinematically allowed.  The resulting cross-section behaves as $\sigma
\sim s^{n/2}/ M^{n+2}$. This expression grows with $s$; if it were
valid for all $s$, it would violate unitarity.
We will see that string theory supplies an appropriate form factor to
cut off this dependence.

%%%%%%%%%%%%%%%%%%%%%%%%%%%%%%%%%%%%%%%%%%%%%%%%%%%%%%%%%%%%%%%%%%%%%%
\begin{figure}
\centerline{\epsfysize=1.50truein \epsfbox{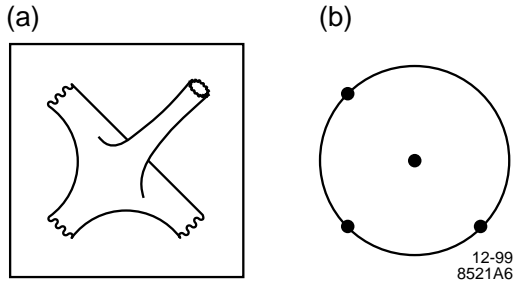}}
\vskip 0.0 cm
\caption{Schematic illustration and world-sheet diagram of the
   scattering process involving three open strings on a D-brane
    and one closed string in the bulk.}
\label{fig:Gemit}
\end{figure}
%%%%%%%%%%%%%%%%%%%%%%%%%%%%%%%%%%%%%%%%%%%%%%%%%%%%%%%%%%%%%%%%%%%%%%

In our stringy toy model, the graviton is a part of the closed string
massless spectrum, while the electrons and photons are described by
massless states of open strings. Therefore, to study the process
$\ee\to \gamma G$ we consider the string scattering amplitude involving
three open strings and a closed string.  The calculation of this
amplitude is very similar to the calculation of the four open-string
scattering presented in Section 3. The amplitude is given by
\beq
\M(1,2,3,G) = g M(1,2,3,G) \,\, \tr ( [t^1, t^2] t^3),
\eeq{Mformula}
where we need to substitute for each $t^i$ the appropriate matrix from
\leqn{threelambdas}. To evaluate the ordered amplitude $M(1,2,3,G)$,
we map the
string worldsheet in Fig.~\ref{fig:Gemit} (a) onto a disc, and then
into the upper half plane. The three open string vertex operators have
to be placed on the boundary; the closed string vertex operator can sit
anywhere inside the upper half plane. Then, the ordered amplitude is
\beq
M(1,2,3,G) = {1 \over {\ap}^2}  X^2
\int_{{\cal C}^+} d^2 z \left<\prod_{i=1}^3 {\cal V}_{q_i} (x_i,k_i) \V_{q_G}
(z, \bar{z},k_G) \right>\ ,
\eeq{3open1closed}
where $\V_{q_i}(x_i)$ is the vertex operator of the open string state
$i$, and $\V_{q_G} (z, \bar{z})$ is the vertex operator of the graviton.
The open string vertex  operators
are placed on the real axis at $x_i = 0, 1, X$, with $X$ to be fixed and
sent to $\infty$.  The integral is taken over the upper half
plane ${\cal C}^+$.
Just as in Section 3, we perform the doubling
trick, extending the definitions of the fields to the full complex
plane; then the open string vertex operators are given by
\leqn{photons} and \leqn{photinos}. The closed string vertex operator
in the 0 picture takes the form
\beqa
\V_{0,0}^{\mu\lambda} (z, \bar{z},k)
 & = & - {\kappa \over \pi \ap}
D^\lambda_{\,\nu}
 \left( \partial X^\mu (z) + i k \cdot \psi (z) \psi^\mu (z)
\right) e^{i k \cdot X(z)} \CR
 && \cdot \left(\partial X^\nu (\bar{z}) + i Dk \cdot \psi (\bar{z})
\psi^\nu (\bar{z}) \right) e^{i Dk \cdot X(\bar{z})},
\eeqa{graviton}
where $D^\mu_{\,\nu}=1$ for $\mu=\nu=0..3$, $D^\mu_{\,\nu}=-1$ for $\mu=\nu=5..
10$, and $D^\mu_{\,\nu}=0$ for $\mu \not= \nu$. Using these vertex operators
and the correlation functions given in \leqn{corr1} and \leqn{corr2}, the
amplitude \leqn{3open1closed} can be evaluated. In this calculation,
one encounters integrals of the form
\beqa
I_0(a,b,c) &=& \int_{{\cal C}^+} d^2z \, |z|^a \, |1-z|^b \, (z-\bar{z})^c, \CR
I_1(a,b,c) &=& \int_{{\cal C}^+} d^2z \, |z|^a \, |1-z|^b \, (z-\bar{z})^c \,
(z + \bar{z}),
\eeqa{masterints}
with arbitrary $a,b,c.$ Using the representation
\beq
|z|^a = {1 \over \G(-a/2)} \int_0^\infty t^{-a/2-1} e^{-t|z|^2} dt
\eeq{za}
these integrals can be evaluated. The results are
\beqa
I_0(a,b,c) &=&  (2i)^c {\sqrt{\pi} \over 2}
\G \left( -1 - (a+b+c)/2 \right) \CR \CR & & \hskip 1.0cm\cdot
{\G \left((1+c)/2\right) \G\left(1+(b+c)/
2\right) \G\left(1+(a+c)/2\right) \over \G(-a/2) \G(-b/2) \G\left(2+(a+b)/
2 + c \right)} \ ; \CR \CR
I_1(a,b,c) &=& 2 {2+a+c \over 4+a+b+2c} I_0 (a,b,c).
\eeqa{masterints1}

We find that the individual amplitudes contributing to \leqn{eegGfield} are
all multiplied by a common factor
\beqa
\F(s,t,u,m^2) &=&  {1 \over \sqrt{\pi}} e^{-(\log 2)\ap m^2} \G(\half - \half
\ap m^2) \CR\CR
& & \hskip -2.0cm
\cdot{\G(1-\half \ap s) \G(1-\half \ap t) \G(1-\half \ap u) \over
\G(1+\half \ap(s-m^2)) \G(1+\half \ap(t-m^2)) \G(1+\half \ap(u-m^2))}\ .
\eeqa{AApref}
An analogous result holds for the process $gg \rightarrow gG$: To obtain
the string theory amplitude, we just multiply the field theory answer by
the same prefactor \leqn{AApref}.   This result is in agreement with
the calculation of Dudas and Mourad \cite{DM}.
We believe, but we have not been able to show, that the relation among
amplitudes
is a consequence of the $N=4$ supersymmetry of the underlying model.
The field-theory cross section formula \leqn{eegGfield} is then modified by
\beq
   {d \sigma\over d \cos\theta}  =
         {d \sigma\over d \cos\theta}\biggr|_{\mbox{\scriptsize ft}} \cdot
\left| \F(s,t,u,m^2) \right|^2 \ .
\eeq{SMGprocesses}

The expression \leqn{AApref} has an interesting pole structure \cite{DM}.
The poles in the $s$ channel occur for $s=2nM_S^2$, and correspond to
producing SR states with an even excitation number. The SR states with an
odd excitation number cannot decay into a graviton and an open string
massless state. On the other hand, these states can mix with the graviton,
leading to the appearance of extra poles at $m^2 = (2n+1) M_S^2$. These
poles were also observed by Hashimoto and Klebanov \cite{Klebanov} in
their calculation of the gluon-gluon-graviton vertex. Their presence is
essential for the correct factorization properties of the form factor
\leqn{AApref}.

The form factor  \leqn{AApref} expresses the way in which
 the amplitudes for KK graviton
emission are cut off in all relevant high-energy limits. Assume that
the kinematic variables are sufficiently far away from
any of the poles in \leqn{AApref}. (Near the poles, the effects of
finite width of the resonances have to be taken into account. This is
beyond the scope of our analysis here.) For the radiation
of state of very high mass, we can evaluate $\F$  at the threshold
$s=m^2$, $t = u = 0$, and then take $m^2$ large. Using Stirling's formula,
we find
\beq
   \F\sim  \exp[-(\log 2) \ap m^2] \ .
\eeq{Ainlargem}
In the limit of fixed mass, $s \to \infty$, and fixed angle, we find
\beq
   \F \sim  \exp[-\ap s f(\cos\theta)] \ ,
\eeq{Ainlarges}
where $f(c)$ is the function defined in \leqn{fthetaform}.  In the
high-energy limit in which $s,t,u, m^2$ all become large together, we
find the more complicated formula
\beq
\F \sim  \exp[-\half \ap s f(x, \cos\theta)] \ ,
\eeq{Ainalllarge}
where $x = m^2/s$ and $f(x,c)$ is given by
\beqa
f(x, c) & = & x\log{4 x} - (1-x){(1+c)\over 2} \log{(1+c)\over 2}
              - (1-x){(1-c)\over 2} \log{(1-c)\over 2}   \CR
      &  & - ({(1+c)\over 2}+ x{(1-c)\over 2})
                        \log  ({(1+c)\over 2}+ x{(1-c)\over 2})\CR
      &  & - ({(1-c)\over 2}+ x{(1+c)\over 2})
                        \log  ({(1-c)\over 2}+ x{(1+c)\over 2})\ .
\eeqa{Ffval}
The function $f(x,c)$ is positive for the allowed values of $c$ and $x$,
even though this property is not manifest in \leqn{Ffval}.
Thus, the string correction  \leqn{AApref}
gives a form factor suppression in all hard-scattering regions.

Recently, Bando \etal\ \cite{Bando} have pointed out that high-mass graviton
emission from a brane is suppressed by a form factor effect due to
brane recoil.  The formula they propose is
\beq
        \F \sim \exp[-\half {\Lambda_S^2\over \tau_3} m^2]  \ ,
\eeq{AfromBando}
where $\tau_3$ is the brane tension and $\Lambda_S$ is a cutoff scale which
should be of order $M_S$.  The expression in the exponent is smaller than
that in \leqn{Ainlargem} by a factor of order $g_{YM}^2$.  In weak-coupling
Type IIB
string theory, brane recoil is described by the emission of scalars in the
$N=4$ gauge multiplet associated with brane.  With the orbifold projection
described in Section 2, there is one scalar $\phi^3$ that survives and
remains in the spectrum. This scalar does not couple to the QED state in
the field theory limit, but it does couple through higher-dimension operators.
However, these couplings are proportional to one factor of $g_{YM}$ in
the amplitude for each $\phi^3$ emitted.  These inelastic processes
deplete the cross section for elastic $G$ emission without $\phi^3$
emission and should lead to a form factor suppression of the form
$\exp[-c g_{YM}^2 m^2/M_S^2]$.  This is in agreement with the result of
\cite{Bando}.  However, we see from \leqn{Ainlargem} that there is a
parametrically more important source for the form factor, the intrinsic
non-pointlike nature of the states in string theory.   We should note that
the numerical coefficient in the formula \leqn{tension} for the brane tension
is quite small, so that effects of the size \leqn{AfromBando} might
nevertheless be relevant.

In our study of open-string scattering,  we saw that the form factor
cutoff of string amplitudes is important only at very high energy.  At
energies of the order of the string scale, a much more important phenomenon
is the enhancement of scattering cross sections through the effect of
SR resonances. We have seen that the amplitudes for
graviton emission contain the series of SR poles at $s = 2n M_S^2$
and $m^2 = (2n+1) M_S^2$.
Thus, string theory predicts an enhancement of the rate for graviton emission
processes such as $\ee\to \gamma G$
through resonant processes such as
\beq
     \ee \to \gamma^{**} \to \gamma G  \ , \quad
   \ee \to \gamma \gamma^{*}_{1,2} \to \gamma \gamma G  \ .
\eeq{Gres}
Typically, the resonances would be seen more clearly in $\ee$ or $q\bar q$
elastic scattering. However, the resonant production of missing-energy
events would be an important confirmation that the observed resonances
were a manifestation of quantum gravity with large extra dimensions.

\section{Stringy corrections to $\gamma\gamma$ scattering}

In this section, we address the question of the relative strengths
of the effective operators in the low energy theory mediated by
virtual SR and KK exchanges. At the end of Section 1, we
argued, on very general grounds, that in any weakly coupled
string theory the SR-mediated operators are expected to dominate.
Here, we will substantiate this claim by an explicit calculation.

\subsection{Tree amplitude}

It is important to note that, unlike renormalizable field theory,
string theory gives a nonzero contribution to the $\gamma\gamma$
scattering amplitude at the tree level.  To compute this amplitude,
we follow the procedure outlined in Section 3. We find
\beq
\A(\gamma_R\gamma_R \to
 \gamma_R\gamma_R) = -  e^2 s^2 \left[ {1\over st}\S(s,t) + {1\over su} \S(s,u)
                    +  {1\over tu} \S(t,u)\right] \ ,
\eeq{twogam}
where $\S(s,t)$ is given by \leqn{Sfactor}.  The helicity amplitudes for
$\gamma_R\gamma_L \to \gamma_R\gamma_L$ and $\gamma_L\gamma_L
\to \gamma_L\gamma_L$ can be obtained from \leqn{twogam}
by crossing.  All other helicity amplitudes vanish.

The expression \leqn{twogam} must vanish in the field theory limit $\ap\to 0$.
This is easily seen as a consequence of $s+t+u = 0$.  Using a higher--order
expansion of $\S$, as in \leqn{primexpand}, we obtain
\beq
\A(\gamma_R\gamma_R \to
 \gamma_R\gamma_R) = {\pi^2\over 2} e^2 {s^2\over M_S^4} + \cdots \ , \qquad
\A(\gamma_R\gamma_L \to
 \gamma_R\gamma_L) = {\pi^2\over 2} e^2 {u^2\over M_S^4} + \cdots \ .
\eeq{twogamlim}
This result can be compared to the $\gamma\gamma\to \gamma\gamma$ amplitude
induced by KK graviton exchange.  Using the effective Lagrangian \leqn{TTint},
it is straightforward to see that \cite{Cheung,Hooman}
\beq
\A_{\rm KK}(\gamma_R\gamma_R \to
 \gamma_R\gamma_R) = 16{\lambda\over M_H^4} s^2 \ , \qquad
\A_{\rm KK}(\gamma_R\gamma_L \to
 \gamma_R\gamma_L) =  16 {\lambda\over M_H^4} u^2 \ .
\eeq{twogamKK}
These expressions have exactly the same form as \leqn{twogamlim}, and this
must be so, because there is only one gauge-invariant, parity-conserving
dimension 8 operator which contributes to $\gamma\gamma\to\gamma\gamma $.
However, the scale $M_H$ in \leqn{twogamKK} is different from the string
scale that appears in \leqn{twogamlim}. We have already remarked in Section
4 that the relation between $M_S$ and $M_H$ can be obtained explicitly in
our string model, and that in a weakly-coupled string theory the effect of
KK graviton exchanges \leqn{twogamKK} is subdominant to the SR exchanges
\leqn{twogamlim}. In the next section, we will derive that
result.

%%%%%%%%%%%%%%%%%%%%%%%%%%%%%%%%%%%%%%%%%%%%%%%%%%%%%%%%%%%%%%%%%%%%%%
\begin{figure}
\centerline{\epsfysize=1.50truein \epsfbox{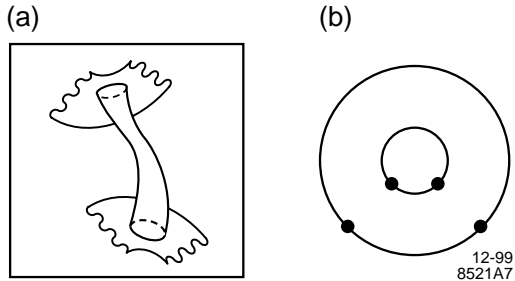}}
\vskip 0.0 cm
\caption{Schematic illustration and world-sheet diagram for
   open string scattering via a closed string exchange.}
\label{fig:lovelace}
\end{figure}
%%%%%%%%%%%%%%%%%%%%%%%%%%%%%%%%%%%%%%%%%%%%%%%%%%%%%%%%%%%%%%%%%%%%%%

\subsection{Loop amplitude}

In string theory, the graviton exchange proper arises at the next order
in perturbation theory.  The graviton is a closed-string state.  It first
appears in open-string perturbation theory through the 1-loop diagram
shown in Figure~\ref{fig:lovelace} \cite{Lovelace}.
In this section, we will compute this diagram and show that it contains
a piece which has the form of the one-graviton exchange amplitude.
Some other properties of this diagram have recently been analyzed in
\cite{DM}.

In the covariant formulation of string theory \cite{bigbook,CFT},
the open string loop amplitude shown in Figure~\ref{fig:lovelace} is
computed in terms of correlation functions of vertex operators placed
on the two boundaries.  It is convenient to conformally map the annulus
shown in Figure~\ref{fig:lovelace}(b) into a cylinder, represented by
a rectangle in the complex plane
\beq
0 \leq \Re w \leq \pi\ , \qquad  0 \leq \Im w \leq 2 \pi t \ ,
\eeq{cyl}
periodically connected with the identification $w \cong w + 2 \pi i t$.
The boundaries of the annulus are mapped to the lines $\Re w = 0$ and
$\Re w = \pi$.  The parameter $t$ is a modulus which must be integrated
over the whole range $0 < t < \infty$.

The complete four-point open string amplitude is a sum of ordered
amplitudes in which the four vertex operators are placed on the boundaries
in all possible ways. The open strings on a D-brane and
the Type IIB closed strings are
oriented, so  we do not need to consider
non-orientable worldsheets such as the  Mobius strip. Thus,
\beqa
\A_{\rm 1-loop} =& g^4 A_{\rm p}(1,2,3,4) \cdot \tr [t^1 t^2 t^3 t^4] +
{\rm \,\,perms\,\,} + \cr\cr
& g^4 A_{\rm np} (1,2;3,4) \cdot \tr[t^1 t^2] \tr[t^3 t^4] +
{\rm \,\,perms}.
\eeqa{1lAformula}
This equation is the analogue of the tree-level color decomposition in
\leqn{Aformula}.  Only the second line, the `non-planar' amplitude, has
the correct color structure to represent graviton exchange.  We will
show that the first term in the second line, which we denote $\A_{Gs}$,
contains the contribution of a virtual graviton exchanged in the $s$-channel.

The explicit expression for $\A_{Gs}$ is
\beqa
{\cal A}_{Gs} = g^4  \tr[t^1 t^2] \tr[t^3 t^4] \int {dt \over t}
 \left[ \prod_{i=1}^4 \int_0^{2\pi t} dy_i\right] \CR
 \cdot   Z_x^p
\sum_\lambda    Z_\lambda \left< \prod_{i=1}^4
 \eps_i\cdot {\cal V}_0 (w_i, k_i)
\right>_\lambda \ ,
\eeqa{loopamp}
where $Z_x^p$ denotes the partition function of the worldsheet bosons $X^\mu$
and the anticommuting ghosts, and $Z_\lambda$ denotes the partition function of the
worldsheet fermions $\psi^\mu$ and the commuting ghosts.  The
expectation value is correspondingly assumed to be computed only from field
contractions, excluding the partition functions.  The parameter  $\lambda$
denotes the periodicities of the worldsheet fermions.  As we stated in
Section 2, we will carry out our computations in this section in the original
$N=4$ supersymmetric Type IIB theory.  Thus, we will sum only over uniform
periodic and antiperiodic boundary conditions for the
world-sheet fermions around each of the two  cycles.
 The vertex
operators are placed at
\beq
w_1 = i y_1 \ , \quad w_2 = i y_2 \ , \quad
w_3 = \pi + i y_3\ , \quad w_4 = \pi + i y_4 \ .
\eeq{positions}
We will check the overall normalization of this expression in Section 6.3.

The easiest way to account for the boundary conditions on the worldsheet
fermions is to extend their definitions
to $\pi \leq \Re w \leq 2\pi$. On this extended worldsheet,
the fermions are holomorphic, and their possible
periodicities and correlators are the same as for a torus with
modulus $it$. For the worldsheet bosons, the boundary conditions
can be described using the method of image charges. For the fields
located on the boundary and satisfying Neumann boundary conditions
the correlator is the same as that for a torus with modulus $it$, with
an extra factor of 2 from the image fields. The correlators necessary
for our calculation are listed explicitly in Appendix B.

For the computation of this section, we will be interested in the contribution
to the amplitude from bosonic closed string states propagating up the cylinder.
These states have fermions antiperiodic around the cylinder, that is, in
the direction of $\Im w$.  Both boundary conditions in the direction of
$\Re w$ are needed to enforce the GSO projection \cite{bigbook}.  We will
refer to the partition functions for the sectors antiperiodic in the
imaginary direction and antiperiodic/periodic in the real direction as
$Z^A_A$/$Z^A_P$ and use a similar notation for the correlation functions.
 In Section 6.3, we will also consider the contribution from
bosonic open string states propagating around the cylinder.  These states
have fermions with boundary conditions antiperiodic in the real direction.
The computation will involve the partition functions $Z^A_A$/$Z^P_A$
and the analogous correlators.

For the cylinder amplitude, the superconformal charges satisfy
 $\sum_i q_i = 0$.  Thus, we will write all four vertex operators in the
0 picture.  We will use the explicit form
\beq
\V_0^\mu (k_i) =  (i \dot{X}^\mu + \ap 2k \cdot \psi \,\,\psi^\mu)
e^{i k_i \cdot X} (w_i, \bar{w_i}),
\eeq{photons1}
where the  dot denotes a derivative along the boundary.
Note that this expression is slightly different from \leqn{photons} in
that the $X$ field has not been split into holomorphic and
antiholomorphic components.

The $t$ integration in \leqn{loopamp} runs from 0 to $\infty$.  However,
this domain of integration can be separated into two regions.  In the
limit of small $t$, the cylinder becomes very long and the amplitude
is dominated by light closed-string states.  In the limit of large $t$,
the cylinder becomes very narrow and the amplitude is dominated by
light open-string states.   The separation between
these two regions is ambiguous, since only their sum is a well-defined
gauge-invariant quantity. We parametrize this ambiguity
by the integration cutoff $t_0$. Below we will show that the
small-$t$ region reproduces the graviton
exchange amplitudes \leqn{twogamKK}, with $M_H$ related to the string
scale and $t_0$. In this calculation, we will use the small-$t$
expansions of the partition functions and correlators.  These
expressions (given in Appendix B) are valid up to $t \sim \pi$.
 This suggests that the natural
value of the cutoff is $t_0 \sim \pi$.  The expression we will derive
for $M_H$ will depend on $t_0$.  This simply makes clear that the
loop diagrams of string
theory also  give other contributions to the dimension 8 terms of the
effective Lagrangian.  The most important point is that all of these
contributions are subleading, suppressed by a power of $g^2$ relative to
the SR contribution  \leqn{twogam}.

We now describe the evaluation of the graviton-exchange contribution in
\leqn{loopamp}.   For the moment, we consider a  D$p$-brane with $p$
 arbitrary;  later we will specialize to the case $p=3$. Using the
small-$t$ expressions of the partition functions and correlators given
in Appendix B, we find the expression
\beqa
{\cal A}_{Gs} & = &g^4 \delta^{12}\delta^{34} \cdot 4^{-\ap s} 2^{(7-3p)/2} \pi^{3-p}
{\ap}^{(7-p)/2} \cdot
\int_0^{t_0} dt   t^{(5-p)/2}   \exp({\ap s \over 2}
{\pi \over t})  \CR && \cdot 2 \left[ \prod_{i=1}^4 \int_0^1 dY_i \right]
(\sin \pi Y_{12})^{-\ap s} (\sin \pi Y_{34})^{-\ap s} F(Y_i; \eps_i, k_i)
\hskip.5cm + \hskip.5cm \Delta,
\eeqa{loopans}
where $F$ is a function of external momenta and kinematics which has
no $t$ dependence, $Y_i = y_i/2\pi t$, $Y_{ij}=Y_i-Y_j$, and $\Delta$ is the
contribution to the integral from the large-$t$ region. Explicitly,
the function $F$ is given by
\beq
F =  C_1 + C_2 \ ,
\eeq{Fisthis}
where
\beqa
C_1 &=& \left( {1 \over 2\ap} \right)^2 \eps_1 \cdot \eps_2
\eps_3 \cdot \eps_4 \sin^{-2} \pi Y_{12} \sin^{-2} \pi Y_{34}, \CR
C_2 &=& {k_1 \cdot k_4} \bigl( 2 k_1 \cdot k_4 \, \eps_1 \cdot \eps_2 \, 
\eps_3 \cdot \eps_4
+2 \eps_1 \cdot \eps_2 \, (k_1 \cdot \eps_3 \, k_3 \cdot \eps_4 +
\eps_3 \cdot k_4 \, k_2 \cdot \eps_4) \CR
& & \hskip 2.0cm
 +2 \eps_3 \cdot \eps_4 \, (k_1 \cdot \eps_2 \, \eps_1 \cdot k_3 +
\eps_1 \cdot k_2 \, \eps_2 \cdot k_4) \bigr).
\eeqa{explicit}
Since we are only interested in the $s \rightarrow 0 $ limit of the
amplitude, in \leqn{explicit} we have dropped the terms which do not
contribute in this limit.

The small-$t$ contribution in \leqn{loopans} factorizes into two
integrals, the modulus intergal in the first line and the coordinate
integral in the second line. The coordinate integral can be easily
evaluated. In this calculation, one encounters two simple integrals,
\beq
I_1 = \int_0^1 dY_1 \int_0^1 dY_2 (\sin \pi Y_{12} )^{-2-\ap s},
\eeq{firstI}
and
\beq
I_2 = \int_0^1 dY_1 \int_0^1 dY_2 (\sin \pi Y_{12} )^{-\ap s}.
\eeq{secondI}
Evaluating these integrals in the limit $\ap s \rightarrow 0$ yields
$I_1=0$, $I_2=1$. Therefore, in this limit we have
\beq
 2 \left[ \prod_{i=1}^4 \int_0^1 dY_i \right]
(\sin \pi Y_{12})^{-\ap s} (\sin \pi Y_{34})^{-\ap s} F(Y_i; \eps_i, k_i)
= 2 C_2.
\eeq{thirdI}
One can show that this expression is identical to the matrix element
of the square of the photon energy-momentum tensor,
$T^{\mu\nu}(1,2)T_{\mu\nu}(3,4)$.
This means that in this limit, this
process is accurately described by the effective Lagrangian \leqn{TTint}.
The integral over the modulus $t$ then determines the coefficient of
this operator.

The modulus integral can be rewritten in a form
reminiscent of a massive graviton propagator from field theory.
To do this, we change the integration variable
to $v=1/t$, and use the identity
\beq
v^{(p-9)/2} = \left( {\ap \over 2} \right)^{(9-p)/2} \int d^{9-p}m
\exp(-{\pi \ap m^2 \over 2} v)\ .
\eeq{thevmid}
Performing the $v$ integration, we find
\beq
\int_0^{t_0} dt   t^{(5-p)/2}   \exp({\ap s \over 2}
{\pi \over t}) = -
\left( {\ap \over 2} \right)^{(7-p)/2} {1 \over \pi} \int d^{9-p}m
{1 \over s-m^2} \exp\bigl({\pi \ap (s-m^2) \over 2} v_0\bigr),
\eeq{likeprop}
where $v_0=1/t_0$. When both $s$ and $m^2$ are small compared to
$1/\ap$, the integrand in \leqn{likeprop} is just the field theory
propagator. We have already pointed out that the virtual graviton
exchange cannot be analyzed within effective field theory; technically,
this results from the divergence of the KK mass integration in the region
of high $m$. The integral in \leqn{likeprop}, however, is finite, due to
the exponential suppression for  $\ap m^2 \gg 1$. This finite coefficient
gives the
scale $M_H$ in \leqn{TTint}.

Evaluating the integral \leqn{likeprop} for $s=0$
and assembling the pieces, we obtain as the leading term in the
low-energy expansion of the small-$t$ integral of  \leqn{loopans}
\beq
{\cal A}_{Gs} =  g^4\delta^{12} \delta^{34} \cdot 2^{(9-3p)/2} \pi^{(13-3p)/2}
(\pi v_0)^{(p-7)/2}
{\ap}^{(7-p)/2} {1 \over 7-p} \cdot
T^{\mu\nu}(1,2) T_{\mu\nu}(3,4) + \ldots \ .
\eeq{stringans}
Now set $p=3$. The  amplitude \leqn{stringans} can
be reproduced by the effective Lagrangian \leqn{TTint}, provided that
we identify
\beq
{8 \over M_H^4} =  g^4 {\pi^2 \over 4} {1 \over M_S^4} \cdot (\pi v_0)^{-2},
\eeq{match1}
and use $\lambda = +1$ in \leqn{TTint}.
As we have explained above, for a
numerical estimate we can evaluate this expression with  $v_0 \sim 1/\pi$.
This gives
\beq
{1 \over M_H^4} \sim  {\pi^2 \over 32}  {g^4 \over M_S^4}.
\eeq{match2}
As expected, the relation is of the form \leqn{MHrel}, with an additional
suppression from the numerical coefficient on the right-hand side.
Substituting this value of $M_H$ into \leqn{twogamKK}, we
confirm that this contribution
 is subdominant with respect to the SR exchange
amplitude \leqn{twogam}.

\subsection{Normalization}

There is another reason that we must analyze the one-loop diagram, and that
is to find the relation between the effective Newton constant or the gravity
scale $M$
and the more fundamental string theory parameters $g$ and $\ap$.
We have already quoted this relation in \leqn{MMSrelation}.  In this
section, we will give the derivation.  Once again, our analysis will be
done for the toy case of an $N=4$ supersymmetric D-brane theory.

\begin{figure}
\centerline{\epsfysize=3.00truein \epsfbox{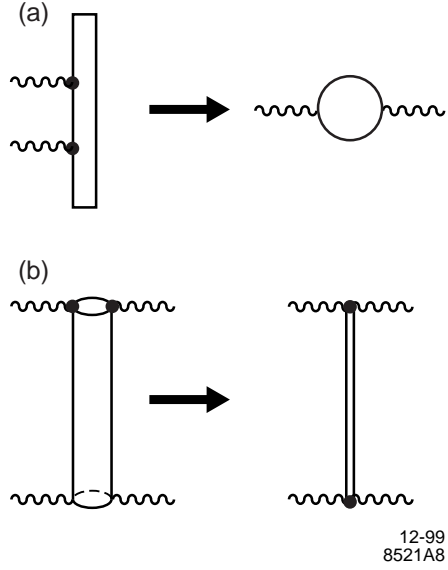}}
\vskip 0.0 cm
\caption{Limits of the cylinder diagram which must be compared to derive
the normalization of the graviton exchange contribution: (a) $t\to \infty$;
(b) $t \to 0$.}
\label{fig:limits}
\end{figure}

Our procedure is illustrated in Figure~\ref{fig:limits}.  We will first
take the $t \to \infty$ limit of the cylinder and relate this to a
loop diagram of Yang-Mills field theory.  This will determine the
normalization of the diagram.  Then we will take the $t\to 0$ limit
to identify the graviton exchange.  In this section, we will give what
we consider the shortest route through this analysis, considering a
two-point function in the first part of the calculation rather than
a four-point function, and, in the second part,
 considering only one fairly simple structure
in the gravitation interaction.

%%%%%%%%%%%%%%%%%%%%%%%%%%%%%%%%%%%%%%%%%%%%%%%%%%%%%%%%%%%%%%%%%%%%%%
%%%%%%%%%%%%%%%%%%%%%%%%%%%%%%%%%%%%%%%%%%%%%%%%%%%%%%%%%%%%%%%%%%%%%%

We thus consider first the $t\to \infty$ limit.
  In principle, we should study the four-point
loop diagram.  However, it is simpler to analyze the two-point function.
The normalizations of these diagrams are related by considering the
limit $(k_1+k_2)^2\to 0$,
 in which pairs of vertex operators factorize into single
vertex operator insertions as shown in Figure~\ref{fig:factorize}.
Through this relation, the normalization of \leqn{loopamp} is equivalent
to the following normalization of the planar two-point loop amplitude
shown in Figure~\ref{fig:limits}(a):
\beq
\A_{2} = g^2  \tr[t^1 t^2] \tr[ 1] \int {dt \over t}
 \left[ \prod_{i=1}^2 \int_0^{2\pi t} dy_i\right]
\cdot   Z_x^p
\sum_\lambda    Z_\lambda \left< \prod_{i=1}^2
 \eps_i\cdot {\cal V}_0 (w_i, k_i)\right>_\lambda \ ,
\eeq{loopamptwo}
where the notations are as in \leqn{loopamp} and the two vertex operators
are placed at $w_1$ and $w_2$ in \leqn{positions}.

It is simplest to concentrate on the structure
\beq
         \epsilon_1 \cdot k_2 \epsilon_2 \cdot k_1 \ .
\eeq{firststructure}
Looking back to the form \leqn{photons1}, we see that this structure
arises in two ways in the contraction of vertex operators, from the
 contraction of  the two factors $\dot X$ with factors $k\cdot X$ in the
exponentials, and from a contraction of the fermionic terms with one
another.  The correlators for $X$ and $\psi$ should be taken in the
limit $t \to \infty$; the appropriate expressions are
given in \leqn{togetherLT}.
In the two sectors corresponding to bosonic open string states,
these terms give
\beqa
  \left< \Pi \epsilon\cdot \V \right>^A_A &\sim&
  \epsilon_1 \cdot k_2 \epsilon_2 \cdot k_1 \left[ {\ap}^2(1 - 2Y)^2
           - 4 {\ap}^2 (e^{-\pi t Y} + e^{-\pi t (1-Y)})^2 \right] \ ,\CR
  \left< \Pi \epsilon\cdot \V \right>^P_A &\sim&
  \epsilon_1 \cdot k_2 \epsilon_2 \cdot k_1 \left[ {\ap}^2(1 - 2Y)^2
           - 4 {\ap}^2 (e^{-\pi t Y} - e^{-\pi t (1-Y)})^2 \right] \ ,
\eeqa{firstVs}
where $Y = Y_{12}$ and, for clarity, we have left off the expectation value
of the exponentials.  Restoring this factor, including the partition
functions from \leqn{partsLT}, and making the cancellations between the
two sectors, we find
\beqa
\A_2 &=& g^2 N_C \delta^{12} \int^\infty_0 {dt\over t}
    {64 \pi^2 t^2 \ap^2\over (8 \pi^2 \ap t)^{d/2}} \CR
& & \hskip 0.5in \cdot \int^1_0 dY \,
 \epsilon_1 \cdot k_2 \epsilon_2 \cdot k_1 \left[ (1 - 2Y)^2 - 1\right]
 \exp[ \ap k_1\cdot k_2 (2\pi t) Y(1-Y)] \ ,
\eeqa{secondVs}
where we have replaced $(p+1) = d$ and $\tr[1] = N_C$.
  Now do the $t$ integral.  For
$d$ close to 4, we obtain,
\beqa
\A_2 &=& g^2 N_C \delta^{12} {8\over (4\pi)^2} \Gamma\left(2-{d\over 2}\right)
 \epsilon_1 \cdot k_2 \epsilon_2 \cdot k_1\,   \int^1_0 dY \,
 \left[ (1 - 2Y)^2 - 1\right] \ .
\eeqa{thirdVs}

As Kaplunovsky \cite{Kaplun} pointed out for the analogous closed string
calculation, this result can be matched to the computation of the one-loop
two-point diagram
in Yang-Mills theory in the background-field gauge. The
required expressions are given in \cite{PskSch}.  The value of this
diagram given there, summed over the bosonic content of the $N=4$ supersymmetric
Yang-Mills theory (1 vector and 6 scalars), is
\beq
 g^2 N_C \delta^{12} {1\over (4\pi)^2 }\Gamma\left(2-{d\over 2}\right)
( \epsilon_1 \cdot \epsilon_2 k_1\cdot k_2 -  \epsilon_1 \cdot k_2
\epsilon_2 \cdot k_1)
 \,   \int^1_0 dY \,
 \left[ 8(1 - 2Y)^2 - 8\right] \ .
\eeq{FTVs}
In this expression, $Y$ is the Feynman parameter.  The first term in the
bracket comes from a spin-independent determinant, the second term from the
spin operator.  The expressions \leqn{thirdVs} and \leqn{FTVs} match.
Thus, the normalization assumed in \leqn{loopamptwo} and in \leqn{loopamp}
is correct.

Now we turn to the $t\to 0$ limit. Here it is simplest to extract the
graviton exchange by considering the limit of high-energy scattering
with low momentum transfer.  That is, we set
\beq
  k_2 \approx -k_1  \ , \qquad   k_4 \approx - k_3  \ .
\eeq{thelimit}
Then the usual graviton exchange diagram in four dimensions contains a
term
\beq
        \A = - 8\pi G_N \delta^{12}\delta^{34}  (2k_1^\mu k_1^\nu) {1\over s}
                             (2k_{3\mu} k_{3\nu})
                     = - 8\pi G_N  {t^2\over s} \delta^{12}\delta^{34} \ ,
\eeq{usualG}
where $s = - (k_1+k_2)^2 =  - (k_3+k_4)^2$ and $t = - (k_1 + k_3)^2$.

In the scattering amplitude of four vector bosons, this term has the
structure
\beq
      \epsilon_1\cdot \epsilon_2 \epsilon_3\cdot \epsilon_4 k_1\cdot k_3
              k_1 \cdot k_3 \ ,
\eeq{secondstructure}
using \leqn{thelimit} to replace $k_2$ and $k_4$.

After close examination
of the various terms contributing to \leqn{loopamp}, one can see that,
after the  cancellation between the $Z^A_A$ and $Z^A_P$ sectors, there is
only one source for a term of this structure.  That is the contribution
in which one takes only the fermionic term in each vertex operator
\leqn{photons1} and contracts the $\epsilon\cdot \psi$ operators on the
same side of the cylinder and the $k\cdot\psi$ operators across the
cylinder.
The correlators needd are given in \leqn{together} and \leqn{across}.
 There are two contractions of this type for each sector.
When these two terms are added, all dependence on the $Y_{ij}$ cancels out.
The contributions from the two sectors then add constructively.  The
sum of these terms gives
\beq
\A_{Gs} = (  \epsilon_1\cdot \epsilon_2 \epsilon_3\cdot \epsilon_4 \delta^{12}
 \delta^{34} t^2) g^4{2 (2\pi\ap)^4\over (8\pi^2 \ap )^{(p+1)/2}}
   \int^\infty_0 dt  t^{(5-p)/2}  \left[\half e^{\pi/4t}\right]^{2\ap s} \ .
\eeq{formofAG}
One should be careful to note that the $t$ in the prefactor is the Mandelstam
invariant, whereas the other factors $t$ represent the modulus of the cylinder.

This expression can be simplified by changing variables from $t$ to $v = 1/t$
and then introducing the variable  $m$ as in \leqn{thevmid}.
Setting also $p=3$, we arrive at the expression
\beq
\A_{Gs} =
( \epsilon_1\cdot \epsilon_2 \epsilon_3\cdot \epsilon_4 \delta^{12}
  \delta^{34} t^2) g^4 {{\ap}^4\over 8\pi}
   \int d^6 m {1\over m^2 -s } \ .
\eeq{next formofAG}
We can  convert the integral over $m$ to a discrete sum over
KK states in the 6 large extra dimensions of periodicity $2\pi R$ by using
the relation
\beq
       R^6 \int d^6 m =   \sum_m  \ .
\eeq{KKsum}

Finally, we may pick off the term in the sum that corresponds to the massless
graviton in four dimensions.  We then identify
\beq
           8 \pi G_N  =  {1\over 8\pi} {\ap}^4 g^4 R^{-6} \ .
\eeq{epG}
Replacing $G_N$ with the fundamental quantum gravity scale $M$ according
to \leqn{fundamental}, we find
\beq
               M^{-8} = \pi \alpha^2 {\ap}^4 \ ,
\eeq{epGnext}
which is equivalent to the promised relation \leqn{MMSrelation}.

\section{Experimental constraints on the quantum gravity scale}

It is useful to compare the constraints on the large extra dimension
scenario that we have obtained in this paper through model-dependent
string effects to more robust, model-inde\-pen\-dent constraints.  In the
introduction, we noted that previous constraints on large extra dimensions
have come from two sources, searches for missing energy due to gravitation
radiation into the extra dimensions, and searches for contact interactions
due to  KK graviton
exchange.  It has become clear in this paper that the possible
contact interactions are model-dependent and may not be of purely gravitational
origin.  So the truly model-independent constraints come only from
missing-energy experiments.

In Table~\ref{thetable}, we summarize the most important present and future
constraints on the quantum gravity scale $M$
 from missing-energy searches.  This table updates the  table
presented in \cite{MPP} and improves upon it in several important respects.

%%%%%%%%%%%%%%%%%%%%%%%%%%%%%%%%%%%%%%%%%%%%%%%%%%%%%%%%%%%%%%%%%%%%%
\begin{table}
\begin{center}
\begin{tabular}{l l | r | r | r }
  Collider   & &  R / M  ($n=2$)  &  R / M  ($n=4$) &   R / M  ($n=6$) \\
           \hline\hline

   Present: &SN1987A   &  $3 \times 10^{-5}/50000$   &
           $1 \times 10^{-9}/1000$
                 & $6 \times 10^{-11} /100  $ \\ \hline
 &LEP 2   &   $4.8\times 10^{-2}$ / 1200
                                       & $1.9\times 10^{-9}$ / 730 &
                              $6.8 \times 10^{-12}$ / 530 \\ \hline
 & Tevatron  &   $5.5 \times 10^{-2}$ / 1140  & $1.4 \times 10^{-9}$ / 860
              & $4.1 \times 10^{-12}$ / 780 \\ \hline \hline
  Future:
% &Tevatron &  $3.9 \times 10^{-2}$ / 1300
%            &  $1.4 \times 10^{-9}$ / 900
%              & $4.0 \times 10^{-12}$ / 810     \\ \hline
&  LC & $1.2\times 10^{-3}$ / 7700  & $1.2\times 10^{-10}$ / 4500 &
                              $6.5 \times 10^{-13}$ / 3100 \\ \hline
 & LHC &  $4.5 \times 10^{-4}$ /12500  & $5.6\times 10^{-11}$ / 7500 &
                              $2.7 \times 10^{-13}$ / 6000 \\ \hline
\end{tabular}
\end{center}
 \caption{Current and future sensitivities to large extra dimensions from
 missing-energy experiments.  All values for colliders are
expressed as 95\% confidence exclusion limits on the size of
 extra dimensions $R$ (in cm) and the effective Planck scale $M$ (in GeV).
  For the analysis of SN1987A, we give probable-confidence limits.}
\label{thetable}
\end{table}
%%%%%%%%%%%%%%%%%%%%%%%%%%%%%%%%%%%%%%%%%%%%%%%%%%%%%%%%%%%%%%%%%%%%%%%%

The
first line of Table~\ref{thetable} gives the constraints obtained
in \cite{us} from the consistency of the observed neutrino flux from the
supernova SN1987A with the predictions of the stellar collapse models. This
analysis puts an upper bound on the rate of energy loss through
graviton emission.   There exist some strong astrophysical bounds on the
scale of quantum gravity---for  example, \cite{Hall}---but these depend on
assumptions about the cosmological scenario.  The constraint from the
supernova is different in character.  Since we have a reasonable understanding
of the composition of a supernova and of the conditions inside its core
during collapse, it is possible to calculate the gravitional radiation
expected in this process in an unambiguous way.  The typical energy of the
emitted gravitons
is well below a TeV, and so the emission rate calculation uses only the
model-independent low-energy limit of the gravitational coupling.
It is argued in \cite{us}
that, though there are uncertainties in the parameters of the supernova core,
the bounds quoted should be accurate to better than a factor of 2.
The bound for the case of two large extra dimensions ($n=2$) is surprisingly
strong and must be taken seriously.  We note that the values given in the
remaining lines of the table are more precise 95\% confidence exclusion
limits available from accelerator experiments.

The second line of the table gives the constraints arising from the process
$\ee\to \gamma +$ (missing) which have
 been announced by the ALEPH collaboration
\cite{ALEPHmg,butnote}.  Similar constraints on anomalous single photon
production have been announced by the other LEP
experiments \cite{DELPHImg,L3mg,OPALmg}.

The third line of the table is derived from a new search for events with
one jet and missing $E_T$ presented by the CDF collaboration
in~\cite{CDFjetET}.  Of the five cuts on missing $E_T$ presented in this
analysis, the analysis based on the cut $E_T > 200$ GeV turns out to give
the best sensitivity.  We have
applied the formulae in \cite{MPP} to convert
the limit on the cross section to the quoted bounds on $M$.  Note that
these bounds are very close to the estimates in the ``Future Tevatron'' line
of \cite{MPP}.

The fourth line of the table gives the reach of a 1 TeV $\ee$ linear collider
as computed in \cite{MPP}.  However, in the fifth line,
the constraints given in Table~\ref{thetable} for the LHC are much stronger.
This is the result of the observation, made in \cite{Jim}, that at the
LHC there is a dramatic improvement in signal/background if one makes a
very hard $E_T$ cut.  It is advantageous to move this cut to as high a value
as the statistics permit.  The results shown here correspond to the analysis
in \cite{MPP} applied to a  cut at $E_T > 1000$ GeV.

For the LHC search, one may worry that the effective field theory used to
obtain the bounds in Table~\ref{thetable} breaks down for the collisions
of the most energetic partons. In Section 5, we have derived the
form factor which describes the modification of the cross sections at
high energies due to string theory effects. We have shown that at very
high energies, this form factor leads to exponential suppression of the
signal cross section. One might expect that the sensitivity of the LHC
searches will be somewhat lowered by this effect. However, it turns out
that for values of the string scale in the  few-TeV range, this effect
does not significantly alter the signal rates at LHC. In fact, we find a
relatively small effect for typical parton-parton center-of-mass energies
and a dramatic enhancement when partons can combine to the SR resonances,
due to processes analogous to  \leqn{Gres} with an excited gluon or quark
intermediate state.  In the situation in which these states are present,
they would also be seen as resonances in the two-jet invariant mass
distribution. We conclude that in
either case, whether the resonances are observed or not, the bounds in the
last line of
Table~\ref{thetable} would not be significantly decreased
 by stringy physics.

\section{Conclusions}

In this paper, we have studied the phenomenology of large extra dimensions
for the situation in which quantum gravity is represented by a weakly-coupled
string theory.  We have found that, in this case, the signatures of large
extra dimensions which have been considered in the literature up to now are
overshadowed by genuine string effects.  The first sign of new physics is
found in string corrections to Standard Model two-body scattering
cross sections, leading to contact interactions due to string resonances and
to the dramatic appearance of these resonances at colliders.  The fact that
these resonances have not yet been observed allows us to put a lower bound
on the string scale of about 1 TeV.  The corresponding limit on the quantum
gravity
scale, $M > 1.6$ TeV, is much stronger than that of any current accelerator
experiment.  The next generation of colliders should probe values of the
string scale up to 5 TeV and values of the quantum gravity scale above 8 TeV.

The motivation for the idea of large extra dimensions in the work of
Arkani-Hamed, Dimopoulos, and Dvali \cite{Nima1} came from the possibility of
a natural relation between the weak interaction scale and the scale of
quantum gravity.  If this possibility is indeed realized, the linear collider
and the LHC will carry out experimental measurements of string physics.
For many
years, physicists have thought of strings as tiny objects and imagined that
we could observe them in experiments only in some distant era.  It seems
now that this era could be close at hand.

\Acknowledgements

We are grateful to Nima Arkani-Hamed for stimulating this investigation,
and to Nicolas Toumbas, who collaborated with us in the early stages of this
work.  We thank Dimitri Bourilkov for a very useful correspondence concerning
 the LEP 2 Bhabha scattering data.
We also thank Tom Banks, Hooman Davoudiasl,
Savas Dimopoulos, Lance Dixon, Ian Hinchliffe,
Ann Nelson, and Alex Pomarol for helpful
discussions and the Institute for Theoretical Physics at UC Santa Barbara
for hospitality.  The work of SC was
 supported in part by the US National Science Foundation
under contract  PHY--9870115; the work of MP and MEP was
 supported by the US Department of
Energy under contract DE--AC03--76SF00515.

\appendix
 \section{Reference formulae for models of contact interactions}

In this appendix, we give the explicit
expressions for the contact-interaction corrections to
Bhabha scattering that are compared in Figures~\ref{fig:Lthree}
and~\ref{fig:NLC}.  We also give the first contact-interaction corrections
to the $\ee\to \gamma\gamma$ and $\gamma\gamma \to \gamma\gamma$ amplitudes.

The unpolarized
cross section formula for Bhabha scattering can be written in the form
\beq
{d \sigma\over d \cos\theta} = {\pi\alpha^2\over 2 s} \left[
  u^2  (|A_{LL}|^2 + |A_{RR}|^2 ) + 2 t^2 |A_{RL,s}|^2  + 2s^2 |A_{RL,t}|^2
        \right]\ ,
\eeq{BH1}
where
\beqa
   A_{LL} &=& {1\over s} + {1\over t} + {(\half-\sstw)^2\over \sstw\cstw}
                 \left( {1\over s-\mz^2} + {1\over t-\mz^2}\right)
                      +  \Delta_{LL} \CR
   A_{RR} &=& {1\over s} + {1\over t} + {\sstw\over \cstw}
                 \left( {1\over s-\mz^2} + {1\over t-\mz^2}\right)
                      +  \Delta_{RR} \CR
   A_{RL,s} &=& {1\over s}  -  {(\half-\sstw)\over \cstw}
                 {1\over s-\mz^2}
                      + \Delta_{RL,s} \CR
   A_{RL,t} &=& {1\over t}  -  {(\half-\sstw)\over \cstw}
                 {1\over t-\mz^2}
                      + \Delta_{RL,t}  \ .
\eeqa{BH2}
For KK graviton exchange parametrized by \leqn{TTint} \cite{Rizzo},
\beqa
   \Delta_{LL}= \Delta_{RR} &=& {\lambda\over \pi\alpha M_H^4} \left[
                         (u + {3\over 4} s) + (u + {3\over 4} t)\right]\CR
   \Delta_{RL,s} &=&  - {\lambda\over \pi\alpha M_H^4}
                         (t + {3\over 4} s)\CR
  \Delta_{RL,s} &=&  - {\lambda\over \pi\alpha M_H^4}
                         (s + {3\over 4} t) \ .
\eeqa{BH3}

For standard dimension-6 contact interactions \cite{ELP},
\beqa
   \Delta_{LL} &=& 2{\eta_{LL}\over \alpha \Lambda^2}\CR
   \Delta_{RR} &=&  2 {\eta_{RR}\over \alpha \Lambda^2}\CR
  \Delta_{RL,s}= \Delta_{RL,t} &=&  {\eta_{RL}\over \alpha \Lambda^2}\ .
\eeqa{BH4}
The VV case corresponds to $\eta_{LL} = \eta_{RR}
= \eta_{RL} = \pm1$.  The  AA case corresponds to $\eta_{LL} = \eta_{RR}
= -\eta_{RL} = \pm1$.

For the string model described in Sections 2 and 3, the corrections are
more easily described by \leqn{SMprocesses}.

The expressions above are written in such a way that they can easily be
pulled apart into cross sections for definite helicity initial and final
states.  At a high-energy linear collider with a polarized $e^-$ beam,
it is possible to resolve ambiguities in the relative contributions of the
various $\Delta_i$.

For completeness, we note also that the amplitude for $\ee\to \gamma\gamma$,
which is given by \leqn{eeggSex} in our string model, takes the following
form with  KK graviton exchange parametrized by \leqn{TTint} \cite{Hooman2}:
\beq
   \A(e^-_Le^+_R \to \gamma_L\gamma_R) =
-2e^2 \sqrt{{u\over t}}\left[ 1 +
 {\lambda\over \pi\alpha M_H^4} ut  \right] \ .
\eeq{eeggTex}
Thus, in this model, we may identify Drell's $\Lambda_\pm$ parameter as
\beq
       \Lambda_\lambda = (\pi \alpha)^{1/4} M_H  \approx  0.39 M_H \ .
\eeq{DisH}

\section{Ingredients needed for the one-loop calculation in Section 6}

The partition functions for the cylinder with modulus $it$, with
fermion periodicities required for our calculation in Section 6, are:
\beqa
Z_x^p &=& (8\pi^2 \ap t)^{-(p+1)/2} \eta(it)^{-8}; \CR
Z^A_A &=&  \left( {\th_{00}(0 \mid it) \over \eta(it)}
\right)^4; \CR
Z^A_P &=& - \left( {\th_{10}(0 \mid it) \over \eta(it)}
\right)^4; \CR
Z^P_A &=& - \left( {\th_{01}(0 \mid it) \over \eta(it)}
\right)^4.
\eeqa{Zs}
Note that the zero-mode integration in the bosonic
partition function, $Z_x$, was performed only in the directions transverse
to the brane. It turns out that this is the only place in the calculation
which depends on $p$. The small-$t$ expansions of the partition functions
which we will use for the calculation in Section 6.2 are
\beqa
Z_x^p &=& (8\pi^2 \ap)^{-(p+1)/2} t^{(7-p)/2} e^{2\pi/3t} +
\ldots; \CR
Z^A_A &=& e^{\pi/3t} (1 + 8 e^{-\pi/t} + \ldots); \CR
Z^A_P &=& -e^{\pi/3t} (1 - 8 e^{-\pi/t} + \ldots).
\eeqa{parts}
In the calculation in Section 6.3, we will make use of the following
large-$t$ expansions:
\beqa
Z_x^p &=& (8 \pi^2 \ap t)^{-(p+1)/2} e^{2\pi t /3} + \ldots; \CR
Z^A_A &=& e^{\pi t/3} (1 + 8 e^{-\pi t} + \ldots); \CR
Z^P_A &=& - e^{\pi t/3} (1 - 8 e^{-\pi t} + \ldots).
\eeqa{partsLT}
Here and below, we only keep the leading terms in the expansions of
bosonic partition functions and correlators. For fermionic quantities,
we keep the first subleading corrections, since in some cases the
leading terms cancel after different sectors are combined.

We will also need the following correlation functions (all of them are
understood to exclude the corresponding partition function):
\beqa
\left< X^\mu (w_i) X^\nu (w_j) \right> &=& g^{\mu\nu} \left( -\ap \log
\left| \th_{11} \left( {w_{ij} \over 2\pi} \mid it \right) \right|^2 + \ap
{(\Im w_{ij})^2 \over 2 \pi t} \right); \CR
\left< \psi^\mu(w_i) \psi^\nu(w_j) \right>^A_A &=& {g^{\mu\nu} \over 2
\pi} {\th_{00} \left( {w_{ij} \over 2\pi} \mid it \right) \over
\th_{11} \left( {w_{ij} \over 2\pi} \mid it \right)} {\partial_\nu \th_{11}
(0\mid it) \over \th_{00} (0\mid it)}; \CR
\left< \psi^\mu(w_i) \psi^\nu(w_j) \right>^A_P &=& {g^{\mu\nu} \over 2
\pi} {\th_{10} \left( {w_{ij} \over 2\pi} \mid it \right) \over
\th_{11} \left( {w_{ij} \over 2\pi} \mid it \right)} {\partial_\nu \th_{11}
(0\mid it) \over \th_{10} (0\mid it)}; \CR
\left< \psi^\mu(w_i) \psi^\nu(w_j) \right>^P_A &=& {g^{\mu\nu} \over 2
\pi} {\th_{01} \left( {w_{ij} \over 2\pi} \mid it \right) \over
\th_{11} \left( {w_{ij} \over 2\pi} \mid it \right)} {\partial_\nu \th_{11}
(0\mid it) \over \th_{01} (0\mid it)},
\eeqa{corrs}
where $w_{ij}=w_i-w_j.$ The fermionic correlators here are just the same
as for a torus with modulus $it$; they are valid for arbitrary $w_i$'s.
On the other hand, the bosonic correlator in the first line is only
valid for the fields that are placed on the boundary and satisfy Neumann
boundary conditions. It differs from a torus correlator by a factor of
2, which correctly takes into account the image charges in this case.
This correlator is sufficient for our present calculation.

The small-$t$ expansions of the correlators \leqn{corrs} depend
on whether the two fields are on the same side of the cylinder or not. We
can write $w_{ij} = \pi \Delta_{ij} + 2\pi i y_{ij}$, where $y_{ij}=y_i-
y_j$, and $\Delta_{ij}=0$ if $i$ and $j$ are on the same side of the cylinder,
and 1 otherwise (this assumes, without loss of generality, that $i>j$.)
The small-$t$ expansions for the case of $\Delta_{ij}=0$ are,
\beqa
\left< X^\mu (w_i) X^\nu (w_j) \right> &=& g^{\mu\nu} \ap  \left( {\pi
\over 2t} - 2 \log 2 + \log t - 2 \log \sin \pi Y_{ij} \right)
\, + \ldots ; \cr\cr
\left< \dot X^\mu (w_i) X^\nu (w_j) \right> &=& i g^{\mu\nu} {\ap \over
t} \cot \pi Y_{ij} \, + \ldots; \cr\cr
\left< \dot X^\mu (w_i) \dot X^\nu (w_j) \right> &=& g^{\mu\nu} {\ap
\over 2t^2} {1 \over \sin^2 \pi Y_{ij}} \, + \ldots;\cr\cr
\left< \psi^\mu(w_i) \psi^\nu(w_j) \right>^A_A &=& -i g^{\mu\nu} { 1\over
2t} {1 \over \sin \pi Y_{ij}} \left( 1 - 4 e^{-\pi/t} \sin^2 \pi Y_{ij}
+\ldots \right); \cr\cr
\left< \psi^\mu(w_i) \psi^\nu(w_j) \right>^A_P &=& -i g^{\mu\nu} {1 \over
2t} {1 \over \sin \pi Y_{ij}} \left( 1 + 4 e^{-\pi/t} \sin^2 \pi Y_{ij}
+\ldots \right),
\eeqa{together}
where $Y_{ij} = y_{ij}/t$. For the case of $\Delta_{ij}=1$ we get:
\beqa
\left< X^\mu (w_i) X^\nu (w_j) \right> &=& g^{\mu\nu} \ap \log t + \ldots;
\cr\cr
\left< \psi^\mu(w_i) \psi^\nu(w_j) \right>^A_A &=& g^{\mu\nu} {2 \over t}
e^{-\pi/2t} \cos \pi Y_{ij} + \ldots;\cr\cr
\left< \psi^\mu(w_i) \psi^\nu(w_j) \right>^A_P &=& - i g^{\mu\nu} {2 \over
t} e^{-\pi/2t} \sin \pi Y_{ij} + \ldots
\eeqa{across}
The other two correlators, $<\dot X X>$ and $<\dot X \dot X>$, are
in this case suppressed by $e^{-\pi/t}$ and do
 not play a role.

For the calculation in Section 6.3, we need the large-$t$ expansions
of the correlators \leqn{corrs}, with the fields on the same side of
the cylinder. These are given by
\beqa
\left< X^\mu (w_i) X^\nu (w_j) \right> &=& - 2 \pi t \ap g^{\mu\nu}
Y_{ij} (1-Y_{ij}); \CR
\left< \dot X^\mu (w_i) X^\nu (w_j) \right> &=& i \ap g^{\mu\nu}
(1 - 2 Y_{ij}); \CR
\left< \dot X^\mu (w_i) \dot X^\nu (w_j) \right> &=& g^{\mu\nu} {\ap
\over \pi t}; \CR
\left< \psi^\mu(w_i) \psi^\nu(w_j) \right>^A_A &=& -i g^{\mu\nu}
\left( e^{-\pi t Y_{ij}} + e^{-\pi t (1-Y_{ij})} \right); \CR
\left< \psi^\mu(w_i) \psi^\nu(w_j) \right>^P_A &=& -i g^{\mu\nu}
\left( e^{-\pi t Y_{ij}} - e^{-\pi t (1-Y_{ij})} \right).
\eeqa{togetherLT}

\if
Given the expansions \leqn{together} and \leqn{across}, the calculation
of the correlation function in \leqn{loopamp} in the region of small $t$
becomes a straightforward task.
We will perform our calculation for $U(1)$ gauge bosons living on
a D$p$-brane; we will then specialize to the case $p=3$, relevant for
our toy model.
For example, the overall prefactor from
the bosonic path integral is,
\beqa
&C_0 = Z_x \left< \prod_{i=1}^4 e^{i k_i \cdot X(w_i)} \right>_{C^2(t)}
\cr\cr
&= 4^{-\ap s} (8 \pi^2 \ap)^{-(p+1)/2} t^{(7-p)/2} e^{2\pi/3t} \exp({\ap s
\over 2} {\pi \over t}) (\sin \pi Y_{12})^{-\ap s} (\sin \pi Y_{34})^{-\ap s},
\eeqa{c0}
where $s=-2k_1 \cdot k_2$ is the usual Mandelstam variable.
Evaluating the rest of the correlators leads to the expression \leqn{loopans}.

We will now discuss the $Y$ integrations in \leqn{loopans}. (The $t$
integration has been discussed in Section 6.2)
We need to evaluate the expression
\beq
2 \left[ \prod_{i=1}^4 \int_0^1 dY_i \right]
(\sin \pi Y_{12})^{-\ap s} (\sin \pi Y_{34})^{-\ap s} F(Y_i; \eps_i, k_i).
\eeq{Yints}
Explicitly, the function $F$ is given by
\beqa
&F =  C_1 + C_2; \cr\cr
&C_1 = \left( {1 \over 2\ap} \right)^2 \eps_1 \cdot \eps_2
\eps_3 \cdot \eps_4 \sin^{-2} \pi Y_{12} \sin^{-2} \pi Y_{34}, \cr\cr
&C_2 = {t \over 2} \bigl( t \eps_1 \cdot \eps_2 \, \eps_3 \cdot \eps_4
+2 \eps_1 \cdot \eps_2 \, (k_1 \cdot \eps_3 \, k_3 \cdot \eps_4 +
\eps_3 \cdot k_4 \, k_2 \cdot \eps_4) \CR
& +2 \eps_3 \cdot \eps_4 \, (k_1 \cdot \eps_2 \, \eps_1 \cdot k_3 +
\eps_1 \cdot k_2 \, \eps_2 \cdot k_4) \bigr).
\eeqa{explicit}
Since we are only interested in the $s$ pole of the amplitude, in
\leqn{explicit} we have dropped the terms which do not contribute to this
pole.

Integrating over the positions of the vertex operators is now reduced to
two simple integrals, which in the limit $\ap s \rightarrow 0$ are
given by
\beqa
&I_1 = \int_0^1 dY_1 \int_0^1 dY_2 (\sin \pi Y_{12} )^{-2-\ap s}
= - \ap s + \ldots,\cr\cr
&I_2 = \int_0^1 dY_1 \int_0^1 dY_2 (\sin \pi Y_{12} )^{-\ap s} = 1 + \ldots
\eeqa{integrals}
The first equation in \leqn{integrals} implies that the $C_1$ term in
\leqn{explicit} does not in fact contribute to the $s$ pole of the
amplitude.

Using \leqn{integrals}, the original expression \leqn{Yints} reduces
to the product of two photon energy-momentum tensors,
$T^{\mu\nu}(1,2) T_{\mu\nu}(3,4)$.

\fi


\begin{thebibliography}{99}
\bibitem{butnote}
Note that in \cite{Jim} a slightly different convention is used in
which $8\pi G_N$ appears on the right-hand side of \leqn{fundamental}.
\bibitem{conventions}
In \cite{Jim}, the coefficient of the operator   $T^{\mu\nu} T_{\mu\nu}$
 is written $4\pi/\Lambda_T^4$.
In \cite{HLZ}, the coefficient of this operator is written $8\pi/(n-2)M_S^4$,
for $n>2$ compact dimensions. All three conventions appear in the subsequent
 literature.  For the processes considered in this paper, matrix elements of
the trace of $T^{\mu\nu}$ are zero, so we need not consider conventions for
the inclusion of that term.
\end{thebibliography}
\end{document}